\documentstyle[prl,aps,epsfig,twoside]{revtex}

\begin{document}
\title{Quantum Computing with Quantum Dots on Quantum Linear Supports}
\author{K. R. Brown, D. A. Lidar,\thanks{
Current address: Department of Chemistry, 80 St. George Street, University
of Toronto, Toronto, Ontario M5S 3H6} and K. B. Whaley}
\address{Department of Chemistry, University of California, Berkeley}
\maketitle

\begin{abstract}
Motivated by the recently demonstrated ability to attach
quantum dots to polymers at well-defined locations, we propose a
condensed phase analog of the ion trap quantum computer: a scheme
for quantum computation using chemically assembled semiconductor
nanocrystals attached to a linear support. The linear support is either a
molecular string (e.g., DNA) or a nanoscale rod. The phonon modes of the linear support
are used as a quantum information bus between the dots. Our scheme offers
greater flexibility in optimizing material parameters than the ion trap
method, but has additional complications. We discuss the relevant physical
parameters, provide a detailed feasibility study, and suggest materials for
which quantum computation may be possible with this approach. We find
that Si is a potentially promising quantum dot material, already allowing a
5-10 qubits quantum computer to operate with an error threshold of
$10^{-3}$.
\end{abstract}

\section{Introduction}

The tremendous excitement following the discovery of fast quantum algorithms 
\cite{Shor:97,Grover:97a} has led to a proliferation of quantum computer
proposals, some of which have already been realized in a rudimentary
fashion. A representative list includes nuclear spins in liquids \cite
{Cory:97a,Gershenfeld:97,Cory:00} and solids \cite{Yamaguchi:99}, trapped
ions \cite{Cirac:95,Wineland:98,James:98,Sorensen:99}, atoms in microwave
cavities \cite{Turchette:95}, atoms in optical lattices \cite{Brennen:99},
atoms in a photonic band gap material \cite{Woldeyohannes:99,Vats:99,Qiao:99}
, quantum dots \cite
{Loss:98,Burkard:99,Hu:99,Zanardi:98,Imamoglu:99,Bandyopadhyay:00,Tanamoto:00,Brun:00,Biolatti:00}, donor atoms in silicon \cite{Kane:98,Berman:2001} and silicon-germanium arrays \cite
{Vrijen:00}, Josephson junctions \cite
{Shnirman:97,Ioffe:99,Mooij:99,Makhlin:00,Blais:00}, electrons
floating on helium \cite
{Platzman:99}, electrons transported in quantum wires \cite
{Ionicioiu:99,Bertoni:00}, quantum optics \cite{Franson:97,Knill:00},
quantum Hall systems \cite{Privman:98}, and anyons \cite{Kitaev:97,Lloyd:00}. For critical reviews of some of
these proposals see \cite{Brandt:98,DiVincenzo:00,Nielsen:book}. To date, no
single system has emerged as a clear leading candidate. Each proposal has
its relative merits and flaws with respect to the goal of finding a system
which is both scalable and fault tolerant \cite{Preskill:97a}, and is at the
same time technically feasible. In this paper we examine the possibility of
making a solid state analog of a scheme originally proposed for the gas
phase, namely trapped ions. One purpose of making such a study is to
undertake a critical assessment of both the benefits and the disadvantages
which arise on translation of an architecture designed for atomic states
coupled by phonons, to the corresponding architecture for condensed phase
qubits. Our proposal uses quantum dots (semiconductor nanocrystals) and
quantum linear supports (polymers or nanorods) in an ultracold environment.
It relies on recent advances in the ability to chemically attach
nanocrystals to polymers in precisely defined locations. Quantum dots are
coupled through quantized vibrations of the linear support that are induced
by off-resonance laser pulses and information is stored in exciton states of
the dots. Internal operations on exciton states are accomplished using Raman
transitions. We provide here a detailed analysis that allows evaluation of
the merits and demerits of a condensed phase rather than gas phase
implementation.

Semiconductor nanostructures are known as ``quantum dots'' (QDs) when their
size is of the order of or less than the bulk-exciton Bohr-radius. In such
``zero-dimensional'' QDs the electron-hole pairs are confined in all three
dimensions and the translational symmetry that holds for bulk semiconductors
is totally lost. As a result of this quantum confinement the energy-level
continuum of the bulk material changes into a discrete level structure. This
structure is very sensitively dependent on the QD radius and shape, crystal
symmetry, relative dielectric constant (compared to the surrounding medium),
surface effects, and defects. This sensitivity can be used to create and
control a wide range of optical effects \cite{Alivisatos:96b}. In general,
the term ``quantum dot'' is used to refer to both ``0-dimensional''
semiconductor structures embedded within or grown on a larger lattice, {\it 
i.e.}, lattice bound, {\it and} to individual, chemically assembled
semiconductor nanocrystals \cite{Jacak:book}. QDs can be created in a larger
crystal structure by confining a two-dimensional electron gas with
electrodes \cite{Maurer:99}, or by making interface fluctuations in quantum
wells \cite{Bonadeo:98a}. A number of promising proposals for quantum
computation have been made using the lattice-bound dots \cite
{Loss:98,Burkard:99,Hu:99,Zanardi:98,Imamoglu:99,Bandyopadhyay:00,Tanamoto:00}. We consider here instead the chemically assembled semiconductor
nanocrystals. In the
remainder of this paper the term QD will therefore be implicitly understood
to refer specifically to chemically assembled nanocrystals.

A large amount of theoretical and experimental information about nanocrystal
QDs exists. Nanocrystals have been studied for their photoluminescence
properties, linear absorption properties, and non-linear spectroscopy, using
a variety of models and techniques\cite
{Efros:82,Brus:84,Ekimov:85,Xia:89,Hu:90,Vahala:90,Sercel:90,Ekimov:93,Takagahara:93,Efros:92,Efros:93,Efros:96,Norris:96,Hybertsen:94,Nirmal:95,Schmidt:96,Banin:98,Xia:99,Li:00,Herron:90,Lippens:90,Grunberg:97,Hill:95,Hill:96,Leung:98,KWSi,Rama:91,Rama:91a,Zunger:94,Wang:96,Rabani:99,Delley:95,Einevoll:92,Laheld:97,Ohfuti:96,Norris:96a,Empedocles:97}. For reviews see, {\it e.g.}, \cite{Ekimov:96,Lannoo:96,Empedocles:99}.
These studies clarified the roles of size-dependence, lattice structure,
surface effects and environment on the exciton spectrum. However, little
attention has been paid so far to the possibility of using nanocrystal QDs
for quantum computing. One reason may be the difficulty of coupling
nanocrystals. Direct interactions between separate dots are small and
difficult to engineer, so that the route to scalability is not obvious. In
the only other study to date that proposed to use nanocrystals for quantum
computing, Brun and Wang considered a model of nanocrystals attached to a
high-Q microsphere and showed that the interaction between QDs can be
achieved by using whispering gallery modes of the microsphere to entangle
individual qubits \cite{Brun:00}. One problem with realization of this model
is that only a few QDs can be placed on each microsphere. Therefore,
scalability would depend on the ability to connect the microspheres by
optical wires.

An exciting route to bypass the coupling problem for quantum dots is
suggested by the recently demonstrated ability to attach QDs to polymers by
chemical methods at well-defined locations \cite{Alivisatos:96}. We show
below that at sufficiently low temperatures, the QD-polymer system has
quantized vibrational modes that can be used to couple electronic
excitations in quantum dots in a controlled and coherent manner. This
``quantum information bus'' concept derives from the ion trap implementation
of quantum computation proposed by Cirac and Zoller \cite{Cirac:95}. Ion
trap schemes take advantage of addressable multilevel ions that are trapped
in harmonic wells. The ions are then coupled through interaction with their
collective vibrational modes \cite{Cirac:95}.\footnote{
In the original Cirac-Zoller proposal \cite{Cirac:95} the ions are coupled
using the motional ground state, but it was shown later that this
requirement can be relaxed \cite{Sorensen:99}.} This scheme can be extended
to any system of multilevel quantum objects bound by coupled quantum
harmonic oscillators. We apply this approach here to a series of nanocrystal
QDs attached to a linear support. The excitonic states of the QD act as
carriers of quantum information which are coupled to the vibrational states
of the linear support. A linear support is a one-dimensional material ({\it 
e.g.}, a stretched polymer or a clamped nanoscale rod) that is connected at
each end to a wall. The support is contained in either a vacuum or a
non-interacting condensed phase matrix such as liquid helium.

The main advantage of using quantum dots rather than ions is the ability
to control the optical properties of quantum dots by varying the size,
shape and 
composition of the dot. On the other hand, a disadvantage is that the
analysis for quantum dots is complicated by the fact that they are complex
composite objects and are not naturally ``clean''. For example, defects and
surface effects can influence the electronic properties \cite{Kuno:97}. Our
model presupposes that nanocrystals which are sufficiently ``clean'' will
ultimately be available, so this puts some severe demands on the
experimentalist.

Section~\ref{overview} gives an outline of the proposal, describing the
basic physics and the formal similarities with the ion trap scheme. Section~\ref{system} describes the physics of the qubits, namely
the electronic states of quantum dots, and the quantum linear support which
provides the information bus between qubits. A summary of the necessary
requirements of the qubit states is given here. In Section~\ref{operations}
we then show how one-qubit and two-qubit operations can be performed in this
system of coupled quantum dots. Section~\ref{feasibility} discusses the {\it 
feasibility} of undertaking quantum logic, with a detailed analysis of the
constraints imposed by decoherence and physical parameters. Quantitative
estimates are made for several specific candidate systems in Section~\ref
{estimates}, followed by conclusions and discussion in Section~\ref
{conclusions}.

\section{Theoretical Overview}

\label{overview}

\begin{figure}[h]
\epsfig{figure=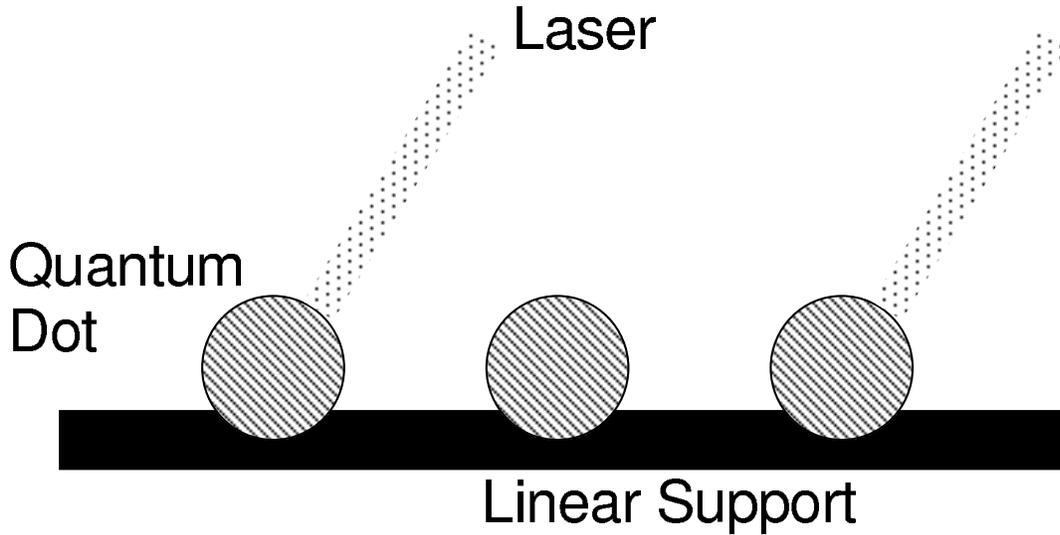,height=200pt}
\caption{Schematic visualization of $N$ quantum dots attached to a linear
support composed of a nano-scale rod or molecular string. Each quantum dot
is addressed by a different laser. The absorption of the dots can be tuned
by varying their sizes, allowing selective addressibility with lasers of
different wavelengths.}
\label{scheme}
\end{figure}

We outline here the basic elements of the quantum dot-quantum linear support
scheme for quantum computation. The proposed system consists of
semiconductor nanocrystal QDs attached at spacings of several tens of
nanometers to a quantum linear support (a string or rod). Each QD supports
one qubit through a certain choice of excitonic states. Single qubit
operations are executed by optical transitions between these states. QDs are
coupled by the linear support in analogy to the ion trap scheme \cite
{Cirac:95}. Thus, one uses detuned laser pulses to excite a phonon of the
quantum linear support, which can then be used to cause conditional
interactions between different dots. The system is depicted schematically in
Figure~\ref{scheme}. The distance between the quantum dots is assumed to be
large relative to their size (see also Section \ref{support} and \ref{approx}.) In the presence of external driving fields, the full Hamiltonian can be
written as the sum of three contributions

\[
H=H_{0}+H_{C}+H_{I}, 
\]
where $H_{0}$ is the free Hamiltonian, $H_{C}$ is the coupling Hamiltonian,
and $H_{I}$ is the Hamiltonian describing the interaction between the system
and the applied laser fields.

The free Hamiltonian, $H_{0}$, is given by 
\begin{equation}
H_{0}=\sum_{n=1}^{N}\sum_{j}\hbar \omega _{nj}^{e}|\Psi _{j}\rangle
_{n}\langle \Psi _{j}|+\sum_{n=1}^{N}\sum_{k}\hbar \omega
_{nk}^{d}b_{nk}^{\dagger }b_{nk}+\sum_{m}\hbar \omega _{m}^{s}a_{m}^{\dagger
}a_{m}+\sum_{l}\hbar \omega _{l}^{f}c_{l}^{\dagger }c_{l}.  \label{eq:zero}
\end{equation}
These four terms represent the energies of the excitons, of the QD phonons,
of the linear support phonons, and of the external electromagnetic field,
respectively. Here $n$ is the QD index, $|\Psi _{j}\rangle _{n}$ is an
exciton eigenstate in the $n^{{\rm th}}$ QD, $b_{nk}$ is an annihilation
operator of the $k^{{\rm th}}$ phonon mode of the $n^{{\rm th}}$\ QD, $a_{m}$
is an annihilation operator of the $m^{{\rm th}}$ linear support phonon
mode, and $c_{l}$ is an annihilation operator of the $l^{{\rm th}}$ mode of
the quantized external electromagnetic field. The phonon frequencies of the
support are denoted by $\omega ^{s}$, and those of the quantum dot by $
\omega ^{d}$.

The coupling Hamiltonian, $H_{C}$, is given by

\begin{equation}
H_{C}=\sum_{njkl}\beta _{njkl}|\Psi _{j}\rangle _{n}\langle \Psi
_{k}|c_{l}^{\dagger }+\sum_{njkl}\gamma _{njik}|\Psi _{j}\rangle _{n}\langle
\Psi _{i}|b_{nk}^{\dagger }+\sum_{njkl}\alpha _{nkl}b_{nk}c_{l}^{\dagger }+ 
{\rm H.c.}  \label{eq:couple}
\end{equation}
The first term is responsible for radiative decay of exciton states. The
most important radiative decay pathway is the recombination of the electron
and the hole. The second term describes the exciton-phonon interaction and
gives rise to both pure dephasing and to non-radiative transitions between
exciton states. The third term is a coupling between the QD phonons and the
electromagnetic field.

The interaction Hamiltonian $H_{I}$ describes the coupling of the excitons
to single-mode plane-wave lasers in a standing wave configuration. We
treat the laser fields semi-classically. A\ QD\ has a permanent dipole
moment due to the different average spatial locations of the electron and
hole. In the dipole limit, we can write

\begin{eqnarray}
H_{I} &=&{\bf D}\cdot {\bf E}  \nonumber \\
&=&{\sum_{{\bf k}ijn}}\;\left[ _{n}\langle \Psi _{i}|e{\cal (}{\bf r}
_{e}^{n}-{\bf r}_{h}^{n})|\Psi _{j}\rangle _{n}|\Psi _{i}\rangle _{n}\langle
\Psi _{j}|\right] \cdot \left[ {\bf \epsilon }_{{\bf k}}{\rm E}_{{\bf k}
}\sin ({\bf k}\cdot {\bf r}_{{\rm cm}}^{n}+\phi _{x}){\rm \cos }(\nu _{{\bf 
k }}t-\phi _{t})\right] .  \label{eq:laser}
\end{eqnarray}
Here ${\bf r}_{e}^{n}$ and ${\bf r}_{h}^{n}$ are the position vectors of the
electron and hole in the $n^{{\rm th}}$ QD respectively; ${\bf r}_{{\rm cm}
}^{n}$ is the center of mass location of this QD; ${\bf \epsilon }_{{\bf k}}$
, ${\rm E}_{{\bf k}}$ and $\nu _{{\bf k}}$ are, respectively, the
polarization, electric field amplitude, and frequency associated with the
field mode ${\bf k}$; $\phi _{x}$ and $\phi _{t}$ are the spatial and
temporal phases of the field. The dipole limit is valid here since a typical
energy scale for single-particle electronic excitations in QDs is $0.1-1$eV,
corresponding to wavelengths $1/k\sim 0.1-1\mu m$. For a typical dot radius $
R\leq 5$ nm, the electric field is then almost homogeneous over the dot.\ In
analogy to ion trap schemes \cite{Cirac:95}, the center of mass of the $n^{
{\rm th}}$ QD, ${\bf r}_{{\rm cm}}^{n}$, is decomposed into its constituent
phonon modes,

\begin{equation}
{\bf r}_{{\rm cm}}^{n}=\sum c_{mn}{\bf q}_{m}=\sum c_{mn}{\bf q}
_{0m}(a_{m}^{\dagger }+a_{m}),  \label{eq:rn}
\end{equation}
where ${\bf q}_{m}$ are normal modes and ${\bf q}_{0m}$ is the zero-point
displacement for the $m^{{\rm th}}$ normal mode, ${\bf q}_{0m}$ =$\sqrt{
\hbar/{2M\omega _{m}}},$ where $M$ is the mass of the mode and $
\omega _{m}$ is the mode frequency. For low phonon occupation numbers, where
the motion of the center of mass of the QD is small compared to the
wavelength of the light, the Lamb-Dicke regime is obtained, {\it i.e.},

\begin{equation}
\eta _{mn{\bf k}}={\bf k}\cdot c_{mn}{\bf q}_{0m}\ll 1.  \label{eq:LD}
\end{equation}
Therefore, we can expand $H_{I}$ to first order in the Lamb-Dicke parameter $
\eta $:

\begin{equation}
H_{I}=2\hbar \sum_{{\bf k}ijn}g_{{\bf k}}^{ijn}|\Psi _{i}\rangle _{n}\langle
\Psi _{j}|{\rm \cos }(\nu _{{\bf k}}t-\phi _{t})\left( \sin \phi
_{x}+\sum_{m}\eta _{mn{\bf k}}(a_{m}^{\dagger }+a_{m}){\rm \cos }\phi
_{x}\right) {\rm .}  \label{eq:Hint}
\end{equation}
Here 
\begin{equation}
g_{{\bf k}}^{ijn}={_{n}}\langle \Psi _{i}|\frac{e{\rm E}_{{\bf k}}}{2\hbar }
\epsilon _{{\bf k}}\cdot {\cal (}{\bf r}_{e}^{n}-{\bf r}_{h}^{n})|\Psi
_{j}\rangle _{n}  \label{eq:g}
\end{equation}
is the resulting coupling parameter between the carrier states in the $n^{
{\rm th}}$ QD. The second term in Eq.~(\ref{eq:Hint}) transfers momentum
from the laser field to the QD, thereby exciting phonon modes of the linear
support. This term allows us to perform two-qubit operations, as described
below. Manipulation of the spatial phase $\phi _{x}$ allows us to
selectively excite either the carrier transition, {\it i.e.}, a change in
the internal degrees of freedom of the QD without changing the vibrational
state of the support, or a side band transition in which the internal
degrees of freedom of both the QD and the vibrational mode of the support
are changed, depending on whether our QD is located at the antinode or node
of the laser, respectively \cite{James:98}.

Now, let $\Omega =2\pi /\tau _{{\rm op}}$ be the Rabi frequency of our
desired quantum operations [see Section~\ref{operations}], and let $T$ be
the temperature. Our system must then satisfy the following set of basic
requirements:

\begin{enumerate}
\item  $\tau _{{\rm op}}<\tau _{{\rm rec}}$, where $\tau _{{\rm rec}}$ is
the time scale for exciton recombination. Typically $\tau _{{\rm rec}
}=10^{-3}-10^{-6}$sec \cite{Efros:96,Heckler:99}.

\item  $\Omega <\omega^s_1$, where $\omega^s_1$ is the first harmonic of the
linear support spectrum. This requirement must be met in order to resolve
the individual support modes.

\item  $k_{b}T<\hbar \omega^s_1$. This ensures that only the ground state
phonon mode is occupied. This requirement comes from the Cirac-Zoller
ion-trap scheme\cite{Cirac:95}, where the motional ground state is used as
the information bus. 

\item  Dephasing and population transfer due to exciton-phonon coupling must
be minimized, or preferably avoided altogether.
\end{enumerate}

We now discuss the details of our system in light of these requirements.

\section{Qubit and Linear Support Definitions}

\label{system}

\subsection{Definition of Qubits}

\label{qubits}

In analogy to the Cirac-Zoller ion trap scheme \cite{Cirac:95}, three
excitonic states will be used, denoted $|0\rangle _{n},|1\rangle _{n},$ and $
|2\rangle _{n}$. Very recent advances in ion trap methodology have allowed
this requirement to be reduced to only two states \cite{Childs:00}
(see also \cite{Monroe:97}). However,
for our purposes it suffices to use the more familiar three state scheme.
The states $|0\rangle _{n}$ and $|1\rangle _{n}$ are the qubit logic states,
and $|2\rangle _{n}$ is an auxiliary state that is used when performing
two-qubit operations. These three exciton states must possess the following
properties:

\begin{enumerate}
\item  Dark for optical recombination:\ This is required so that we will
have long recombination lifetimes.

\item  Dark for radiative relaxation to other exciton states: This is
required to prevent leakage to other exciton states.

\item  Degenerate: The energy separation is required to be smaller than the
lowest energy internal phonon, in order to suppress nonradiative transitions
between states. We wish to make transitions between vibronic eigenstates,
rather than creating oscillating wavepackets which would dephase as they
move on different potential surfaces. Obtaining a large amplitude for moving
between vibrational eigenstates of two surfaces depends on the existence of
two features (i) large Frank-Condon overlap \cite{Haken:book} between these
eigenstates, which will be the case for degenerate exciton potential energy
surfaces; and (ii) very narrow-bandwidth laser pulses which can selectively
address the required states. These transitions are described in Section \ref
{single}. The degeneracy will need to be broken in order to perform certain
operations.
\end{enumerate}

In order to choose states which satisfy the above requirements, detailed
calculation of the exciton wavefunctions and fine structure of the quantum
dots is essential. We employ here the multi-band effective mass model which
has been employed by a number of groups for calculation of the band-edge
exciton fine structure in semiconductor QDs made of direct band gap
semiconductors \cite{Xia:89,Efros:96} . For larger nanocrystals, possessing
radii $R>20$\AA ,~multi-band effective mass theory is generally in
reasonably good agreement with experiment as far as energetics are concerned 
\cite{Leung:98}. It has been used extensively for CdSe nanocrystals by Efros
and co-workers \cite{Norris:96}. While the effective mass approximation
(EMA) has known serious limitations \cite{Lannoo:96}, and has been shown not
to provide quantitative results for smaller nanocrystals \cite{Leung:98}, it
nevertheless provides a convenient, analytically tractable description, with
well defined quantum numbers for individual states, and will allow us to
perform an order of magnitude assessment of the feasibility of our scheme.

To explain the exciton state classification resulting from the multi-band
EMA, it is necessary to consider a hierarchy of physical effects leading to
an assignment of appropriate quantum numbers. These effects are, in
decreasing order of importance:\ (i) quantum confinement (dot of finite
radius, typically smaller than the bulk exciton radius), (ii) discrete
lattice structure (iii)\ spin-orbit coupling, (iv) non-spherical nanocrystal
geometry, and facetting of surfaces, (v) lattice anisotropy ({\it e.g.},
hexagonal lattice), (vi) exchange coupling between electron and hole spin.
The electron-hole Coulomb interaction is neglected: detailed calculations
show that this may be treated perturbatively over the range of nanocrystal
sizes for which the EMA is accurate \cite{Zunger:94,KLeung:97}. These
effects lead to the following set of quantum numbers:\ $n_{e}$ ($n_{h}$) the
principle electron (hole) quantum number, $J_{e}$ ($J_{h})$ the total
electron (hole) angular momentum, $L_{e}$ ($L_{h})$ the lowest angular
momentum of the electron (hole) envelope wavefunction, $S_{e}$ ($S_{h})$ the
electron (hole) Bloch total angular momentum, and the total angular momentum
projection 
\begin{equation}
F_{z}=m_{K}+m_{s},  \label{eq:F}
\end{equation}
where $m_{K}=\pm 1/2,\pm 3/2,\pm 5/2,..$ refers to the projection of the
hole total angular momentum $J_{h}$, and $m_{s}=\pm 1/2,\pm 3/2,\pm 5/2,..$
is the projection of the electron total angular momentum $J_{e}$. State
multiplets are classified by $n_{e}L_{eJ_{e}}$ $n_{h}L_{hJ_{h}},$ e.g. $
1S_{1/2}1P_{3/2}$, and states within the multiplet are labeled by $F_{z}$.
For a II-VI semiconductor such as CdSe, the Bloch states for the valence
band hole states possess total angular momentum $S_{h}=3/2,1/2$, deriving
from coupling of the local orbital angular momentum $1$ in $p$-orbitals,
with hole spin $1/2$. The corresponding Bloch states for the conduction band
electron states have total angular momentum $S_{e}=1/2$, deriving from
coupling of the local orbital angular momentum $0$ in $s$-orbitals, with
electron spin-$1/2$. We consider here only states within the band edge
multiplet, for which $L_{e}=L_{h}=0$, and $S_{e}=1/2,S_{h}=3/2$. Hence the
total electron and hole angular momenta are given by $J_{e}=1/2,J_{h}=3/2$,
respectively and there are a total of eight states within this multiplet. It
follows from Eq.~(\ref{eq:F}) that there is one $F_{z}=2$ state, two $
F_{z}=1 $ states, two $F_{z}=0$ states, two $F_{z}=-1$ states, and one $
F_{z}=-2$ state in this $1S_{1/2}1S_{3/2}$ multiplet. States within a
doublet are distinguished by a superscript ($L$ or $U$). The eigenfunctions,
linear absorption spectrum and selection rules for dipole transitions from
the ground state to this lowest lying EMA multiplet are calculated in Ref. 
\cite{Efros:96}. For spherical QDs the following results were found:

\begin{itemize}
\item  Hexagonal crystal structure: The $F_{z}=\pm 2$ states constitute
degenerate exciton ground states. The $F_{z}=\pm 2$ states and one of the $
F_{z}=0$ states (denoted $0^{L}$) are optically dark in the dipole
approximation.

\item  Cubic crystal structure: The $F_{z}=0^{L},\pm 1^{L},\pm 2$ states
constitute degenerate exciton ground states, and are all optically dark.
\end{itemize}

We consider here explicitly a nanocrystal made from a direct band gap
material with cubic crystal structure. An exciton wavefunction of the $
1S_{1/2}1S_{3/2}$ multiplet, $\Psi _{F_{z}}({\bf r}_{e},{\bf r}_{h})$, can
be expanded in terms of products of single-particle wavefunctions $\psi
_{1/2,m_{s}}^{S}({\bf r}_{e})$ and $\psi _{3/2,m_{K}}^{S}({\bf r}_{h})$~\cite
{Xia:89,Efros:96}. In order to satisfy the requirement of optically dark
qubits (recall condition 1 above), we construct our qubit from the $
F_{z}=-2,0^{L}$ states: 
\begin{eqnarray}
|0\rangle &\equiv &|\Psi _{0^{L}}({\bf r}_{e},{\bf r}_{h})\rangle =\frac{1}{
\sqrt{2}}\left[ |\psi _{1/2,-1/2}^{S}({\bf r}_{e})\psi _{1/2,+1/2}^{S}({\bf 
r }_{h})\rangle -|\psi _{1/2,+1/2}^{S}({\bf r}_{e})\psi _{3/2,-1/2}^{S}({\bf 
r} _{h})\rangle \right]  \nonumber \\
|1\rangle &\equiv &|\Psi _{-2}({\bf r}_{e},{\bf r}_{h})\rangle =|\psi
_{1/2,-1/2}^{S}({\bf r}_{e})\psi _{3/2,-3/2}^{S}({\bf r}_{h})\rangle .
\label{eq:|1>}
\end{eqnarray}
The auxiliary level for cycling transitions is taken to be the $F_{z}=2$
state:

\begin{equation}
|2\rangle \equiv |\Psi _{+2}({\bf r}_{e},{\bf r}_{h})\rangle =|\psi
_{1/2,+1/2}^{S}({\bf r}_{e})\psi _{3/2,+3/2}^{S}({\bf r}_{h})\rangle .
\label{eq:|2>}
\end{equation}
Explicit expressions for the electron and hole wavefunctions are given in
Appendix~\ref{app:Efrosstates}. The states $|0\rangle ,|1\rangle ,|2\rangle $
are degenerate and have equal parity [determined by $(-1)^{F_{z}}$]. 

Naturally, for nonspherical, noncubic, and/or indirect band gap materials,
other states may be more appropriate. It is only essential that they
satisfy the requirements above. In this paper we shall use primarily the
EMA states described above for cubic nanocrystals of direct gap materials,
because they illuminate in an intuitive and quantifiable manner the
difficulties associated with our proposal. However, in the discussion of
feasibility (Section~\ref{feasibility}), we will also present results
obtained with qubit states obtained from tight binding calculations for
nanocrystals constructed from an indirect band gap material (silicon).

\subsection{Quantum Linear Support}

\label{support}

In order to determine whether quantum computation is possible on such a
system we need to examine also the properties of the linear support. The
support is made out of $K$ small units, {\it e.g.}, unit cells or monomers.
We can write the displacement of each unit as a sum of normal modes, 
\begin{equation}
{\bf x}_{k}=\sum_{m}\tilde{c}_{mk}{\bf q}_{m}=\sum_{m}\tilde{c}_{mk}{\bf q}
_{0m}(a_{m}^{\dagger }+a_{m}).  \label{eq:xk}
\end{equation}
The zero point displacements for a homogeneous support are 
\begin{equation}
{\bf q}_{0m}=\sqrt{\hbar /\left( 2\lambda l \omega _{m}\right) },
\label{eq:q0m}
\end{equation}
where $\lambda $ is the linear mass density and $l$ is the length of
the unit. The lowest energy modes will be
long-wavelength transverse modes. Since the wavelengths of the modes of
interest are large compared to the separation between neighboring units, we
can approximate the support as being continuous.

In many cases, a sparse number of attached QDs will have only a small effect
on the normal modes of the support. The validity of this assumption depends
on the materials chosen and will be discussed more thoroughly below. For now
we will calculate all of the relevant properties assuming point-like,
massless quantum dots, consistent with our assumption that the spacing
between the dots is larger relative to their intrinsic size. For the $n^{
{\rm th}}$ point-like QD attached to unit cell $k$, with one dot per unit
cell, we can\ identify the dot and cell normal modes expansion coefficients.
We then have $c_{mn}\equiv \tilde{c}_{mk}$, where $\tilde{c}_{mk}$ and $
c_{mn}$ are, respectively, the coefficients relating the displacement of the 
$k^{{\rm th}}$ unit cell and $n^{{\rm th}}$ QD to the displacement of the $
m^{{\rm th}}$ normal mode [Eqs.~(\ref{eq:rn}) and (\ref{eq:xk}),
respectively]. For a continuous support, the set of $\tilde{c}_{mk}$ becomes
a function $\tilde{c}_{m}(x)$ that is the normalized solution to the wave
equation on the support. Any specific value of $c_{mn}$ can now be written $
c_{m}(x_{n})$, where $x_{n}$ is the position of the $n^{{\rm th}}$ QD.

The two most common types of linear continuous systems are strings and rods.

\subsubsection{Strings}

In a string, the resistance to transverse motion comes from an applied
tension, $\vartheta $. The dispersion relation for the frequency of a string
in mode $m$ is 
\[
\omega _{m}^{s}=\sqrt{\frac{\vartheta }{\lambda }}k_{m}, 
\]
where $\lambda $ is the linear mass density and $k_{m}$ is the wavenumber.
The normalized solution to the transverse wave equation with fixed ends is
given by 
\[
c_{m}(x)=\sqrt{\frac{2l}{L}}\sin (k_{m}x), 
\]
where $k_{m}$ = $m\pi /L$, $l$ is the unit length, and $L$ is the string length.

\subsubsection{Rods}

In a rod, the resistance to transverse motion results from internal forces.
This leads to a different dispersion relation and, consequently, to a
different solution $c_{m}(x)$. The transverse modes of a rod can be defined
in terms of the length $L$, density $\rho $, the Young's modulus $Y$, the
cross-sectional area $A$, and the second moment of $A$ (or the massless
moment of inertia of a slice of the rod), ${\bf I}$. As shown by Nishiguchi 
{\it et al}. \cite{Nishi}, the long wavelength phonon modes ($\lambda \geq
1000$ \AA ) are well described by the classical dispersion equation 
\[
\omega _{m}^{s}=k_{m}^{2}\sqrt{\frac{YI_{z}}{\lambda A}}. 
\]
where $I_{z}$ is the moment in the direction of the displacement.

The transverse normal modes for a clamped rod are well known \cite
{Landau:Elastic}, resulting in the solution 
\begin{eqnarray*}
c_{m}(x) &=&N_{m}\left[ \sin (k_{m}L)-\sinh (k_{m}L)\right] \left[ \cos
(k_{m}x)-\cosh (k_{m}x)\right] \\
&-&\left[ \cos (k_{m}x)-\cosh (k_{m}x)\right] \left[ \sin (k_{m}L)-\sinh
(k_{m}L)\right] .
\end{eqnarray*}
Here $N_{m}$ is a normalization constant proportional to $\sqrt{1/L}$. The
values of $k_{m}$ are not known analytically, but can be shown to be
proportional to $1/L$.

\subsection{Approximations}

\label{approx}

\begin{figure}[h]
\epsfig{figure=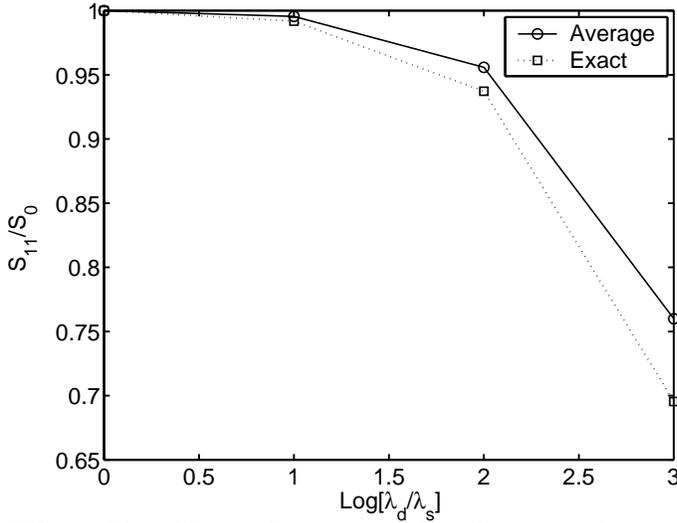,height=200pt}
\caption{ The addition of a sparse number of quantum dots to the linear
support has relatively little effect on the dot modal displacement $S_{nm}$.
Here we present results for $S_{11}$, the dot displacement resulting from
the first harmonic of the support, for a system with two QDs attached to a
string of length $L=2000$ nm. The QDs are centered at 499 and 1501 nm. Each
QD is represented as an increased density that is distributed over a length
of 2 nm, e.g., 499 $\pm 2$ nm. Each dot experiences a displacement which is
affected by the addition of the second dot on the support. The dot
displacement measured relative to the value obtained from a homogeneous
string ($S_{0}$), is plotted as a function of $\log ((\protect\lambda _{d}+
\protect\lambda _{s})/\protect\lambda _{s}$), where $\protect\lambda _{d}$
is the linear density increment due to the dot and $\protect\lambda _{s}$
the linear density of the string. The solid line guides the eye through the
exact solution, the dotted line through the solutions obtained for a
homogenous density equal to the average of $\protect\lambda _{d}$ and $
\protect\lambda _{s}$. One sees that the value of $S_{11}$ changes by less
than a factor of 2 over three orders of magnitude change in the normalized
average density $(\protect\lambda _{d}+\protect\lambda _{s})/\protect\lambda
_{s}$.}
\label{mass}
\end{figure}

The important parameters characterizing the support are $\omega _{1}^{s}$,
the frequency of its first harmonic, and the product 
\begin{equation}
{\bf S}_{nm}\equiv c_{nm}{\bf q}_{0m}.  \label{eq:S}
\end{equation}
This product is the quantum dot displacement resulting from the zero point
motion of mode $m$ of the support. We shall refer to it as the {\em dot
modal displacement}. The above discussion of vibrations in the support
assumed massless QDs, motivated by the assumption that they have negligible
spatial extent relative to the distance between them. To investigate the
effect of the finite mass of the quantum dots, we computed numerical
solutions of the coupled vibration equations for strings and rods having
finite mass increments located at discrete points, simulating the attachment
of finite mass quantum dots. These numerical calculations show that for
sparsely spaced dots of mass small enough that the total weight is the same
order of magnitude as the weight of the support alone, the resulting value
of $S_{nm}$ remains unaffected to within a factor of $2$ by the presence of
the dots (see Figure~\ref{mass}). A simple way to approximate the presence
of the QDs and retain an analytic solution is then to replace the linear
density of the support by the average combined linear density of QD and
support.

Since we are interested here in order of magnitude estimates of feasibility,
we will approximate $c_{m}(x_{n})$ by $\sqrt{2l/L}.$ This approximation
corresponds to the maximum value of $c_{m}(x_{n})$ for a string, and to
approximately the maximum value of $c_{m}(x_{n})$ for a rod.\ Since the
larger the dot displacement, the larger the coupling between dots, this
means that our estimations of number of operations will be an upper bound.

These approximations combined with Eq.~(\ref{eq:LD}) yield the following
equation for the Lamb-Dicke parameter,

\begin{equation}
\eta _{m^{\prime }n{\bf k}_{2}}={\bf k}_{2}\cdot {\bf S}_{nm}=k_{2}S_{nm}
\cos \theta =k_{2}\sqrt{\frac{\hbar }{M\omega _{m}}}\cos \theta ,
\label{eq:LD2}
\end{equation}
where $\theta $ is the angle between the modal displacement and the
direction of the laser beam, and $M$ is the total support mass: $M=L\lambda$.
Note the inverse power dependence on $M$ in Eq.~(\ref{eq:LD2}). We
shall see in 
Section \ref{Parameters} that the massive nature of the linear support and
the resulting small value of the Lamb-Dicke parameter provides the major
limitation for our system.

\section{Qubit Operations}

\label{operations}

\subsection{One-qubit Operations:\ Coupling of Dots to Light}

\label{single}

\subsubsection{Derivation of the Interaction Hamiltonian in the Rotating
Wave Approximation}

While dipole transitions in QDs are similar in principle to dipole
transitions in atomic systems, the strong coupling to internal phonon modes
adds an additional complexity. Consider the modifications of Eqs. (\ref
{eq:zero})-(\ref{eq:laser}) for a single QD {\em unattached} to a linear
support, interacting with a single laser field, with the QD located at the
anti-node of the field, {\it i.e.}, with the $\sin ({\bf k}\cdot {\bf r}_{
{\rm cm}}^{n}+\phi _{x})$ term in Eq.~(\ref{eq:laser})\ vanishing. Thus,
omitting the linear support term: 
\[
H=\sum_{j}\hbar \omega _{j}^{e}|\Psi _{j}\rangle \langle \Psi
_{j}|+\sum_{k}\hbar \omega _{k}^{d}b_{k}^{\dagger }b_{k}+\sum_{ijk}\hbar
\gamma _{jik}|\Psi _{j}\rangle \langle \Psi _{i}|\left( b_{k}^{\dagger
}+b_{k}\right) +\sum_{ij}2\hbar g_{{\bf k}}^{ij}|\Psi _{i}\rangle \langle
\Psi _{j}|\cos (\nu _{{\bf k}}t-\phi _{t}). 
\]
We separate this Hamiltonian into two parts, $H_{0}$ and $H_{I}:$

\begin{eqnarray}
H_{0} &=&\sum_{j}\hbar \omega _{j}^{e}|\Psi _{j}\rangle \langle \Psi
_{j}|+\sum_{k}\hbar \omega _{k}^{d}b_{k}^{\dagger }b_{k}+\sum_{jk}\hbar
\gamma _{jjk}|\Psi _{j}\rangle \langle \Psi _{j}|(b_{k}^{\dagger }+b_{k})
\label{eq:H0} \\
H_{I} &=&\sum_{i\neq j,k}\hbar \gamma _{jik}|\Psi _{j}\rangle \langle \Psi
_{i}|(b_{k}^{\dagger }+b_{k})+\sum_{ij}2\hbar g_{{\bf k}}^{ij}|\Psi
_{i}\rangle \langle \Psi _{j}|\cos (\nu _{{\bf k}}t-\phi _{t})  \label{eq:HI}
\end{eqnarray}
The ``free''\ Hamiltonian $H_{0}$ may be diagonalized by a displacement
transformation. Let $D_{k}(\alpha )$ be the unitary displacement operator: 
\[
D_{k}(\alpha )\equiv e^{\alpha b_{k}^{\dagger }-\alpha ^{\ast
}b_{k}}=D_{k}(-\alpha )^{\dagger }. 
\]
The displaced phonon operator $d_{jk}$ is defined as 
\[
d_{jk}\equiv D_{k}(-\alpha _{jk})b_{k}D_{k}(\alpha _{jk})=b_{k}+\alpha
_{jk}, 
\]
and satisfies standard boson commutation relations: 
\begin{eqnarray*}
\lbrack d_{jk},d_{j^{\prime }k^{\prime }}^{\dagger }] &=&\delta _{kk^{\prime
}} \\
\lbrack d_{jk},d_{j^{\prime }k^{\prime }}] &=&[d_{jk}^{\dagger
},d_{j^{\prime }k^{\prime }}^{\dagger }]=0.
\end{eqnarray*}
Note that for real $\alpha _{jk}$ we have $\alpha _{jk}(b_{k}^{\dagger
}+b_{k})=d_{jk}^{\dagger }d_{jk}-\alpha _{jk}{}^{2}-b_{k}^{\dagger }b_{k}$.
Letting $\alpha _{jk}\equiv \gamma _{jjk}/\omega _{k}^{d}$ and inserting a
complete set of exciton states into Eq.~(\ref{eq:H0}), we find:

\begin{eqnarray*}
H_{0} &=&\sum_{j}\hbar \omega _{j}^{e}|\Psi _{j}\rangle \langle \Psi
_{j}|+\sum_{k}\hbar \omega _{k}^{d}\sum_{j}|\Psi _{j}\rangle \langle \Psi
_{j}|\left( b_{k}^{\dagger }b_{k}-\alpha _{jk}(b_{k}^{\dagger }+b_{k})\right)
\\
&=&\sum_{j}\hbar \omega _{j}^{e}|\Psi _{j}\rangle \langle \Psi
_{j}|+\sum_{jk}\hbar \omega _{k}^{d}|\Psi _{j}\rangle \langle \Psi
_{j}|\left( d_{jk}^{\dagger }d_{jk}-\left( \frac{\gamma _{jjk}}{\omega
_{k}^{d}}\right) ^{2}\right) .
\end{eqnarray*}
The eigenstates of $d_{jk}^{\dagger }d_{jk}$ are labeled $|n_{jk}\rangle $
, where $d_{jk}^{\dagger }d_{jk}|n_{jk}\rangle =$\ $n|n_{jk}\rangle $ and $
\langle n_{jk}|m_{j^{\prime }k^{\prime }}\rangle =\delta _{kk^{\prime
}}F_{nmk}^{jj^{\prime }}$.\footnote{
Note that $n$ refers here to the occupation quantum number of the internal
phonon modes, not to the quantum dot index.} The $F_{nmk}^{jj^{\prime }}$
are Franck-Condon factors \cite{Haken:book}, describing the overlap of
vibrational eigenstates between different excitonic states $j$ and $
j^{\prime }$. We can then rewrite $H_{0}$ as

\[
H_{0}=\sum_{j}\hbar \omega _{j}^{e+}|\Psi _{j}\rangle \langle \Psi
_{j}|+\sum_{jk}\hbar \omega _{k}^{d}|\Psi _{j}\rangle \langle \Psi
_{j}|d_{jk}^{\dagger }d_{jk}, 
\]
where $\omega _{j}^{e+}=\omega _{j}^{e}-\sum_{k}\left( \gamma _{jjk}/\omega
_{k}^{d}\right) ^{2}$ is the renormalized electronic energy level.

We transform to the interaction picture defined by $H_{0}$: $\widetilde{H}
_{I}=\exp \left( iH_{0}t/\hbar \right) H_{I}\exp \left( -iH_{0}t/\hbar
\right) $. To do so, it is useful to insert into this expression two
complete sets of displaced oscillator states belonging to different
excitonic states $i$ and $j$: 
\[
I_{i}\times I_{j}=\bigoplus_{k}\sum_{n}|n_{ik}\rangle \langle n_{ik}|\times
\bigoplus_{k^{\prime }}\sum_{m}|m_{jk^{\prime }}\rangle \langle
m_{jk^{\prime }}|=\bigoplus_{k}\sum_{nm}|n_{ik}\rangle \langle
m_{jk}|F_{nmk}^{ij}. 
\]
Changing variables from $b_{k}$ to $d_{jk}$ in Eq.~(\ref{eq:HI}), and
transforming to the interaction picture now yields, after some standard
algebra:

\begin{eqnarray}
\widetilde{H}_{I} &=&e^{iH_{0}t/\hbar }H_{I}e^{-iH_{0}t/\hbar }  \nonumber \\
&=&\sum_{i\neq j,l}\hbar \gamma _{jik}|\Psi _{i}\rangle \langle \Psi
_{j}|e^{i\omega _{ij}^{e+}t}\bigotimes_{k\neq
l}\sum_{n,m}F_{nmk}^{ij}|n_{ik}\rangle \langle m_{jk}|e^{i\omega
_{k}^{d}(n-m)t}  \nonumber \\
&&\bigotimes_{l}\sum_{n^{\prime },m^{\prime }}F_{n^{\prime }m^{\prime
}l}^{ij}\left( -\frac{\gamma _{jjl}}{\omega _{l}^{d}}|n_{il}^{\prime
}\rangle \langle m_{jl}^{\prime }|e^{i\omega _{l}^{d}(n^{\prime }-m^{\prime
})t}+\sqrt{m^{\prime }+1}|n_{il}^{\prime }\rangle \langle (m^{\prime
}+1)_{jl}|e^{i\omega _{l}^{d}(n^{\prime }-m^{\prime }-1)t}\right)  \nonumber
\\
&&+\sum_{ij}\hbar g_{{\bf k}}^{ij}|\Psi _{i}\rangle \langle \Psi
_{j}|e^{-i(\nu _{{\bf k}}t-\omega _{ij}^{e+}t-\phi
_{t})}\bigotimes_{k}\sum_{n,m}F_{nmk}^{ij}|n_{ik}\rangle \langle
m_{jk}|e^{i\omega _{k}^{d}(n-m)t}+{\rm H.c.}  \label{eq:dip}
\end{eqnarray}
where $\omega _{ij}^{e+}\equiv \omega _{i}^{e+}-\omega _{j}^{e+}$.

While this expression appears very complicated, it can be drastically
simplified under certain reasonable assumptions. First, note that for
single-qubit operations we need to consider only two exciton states $|\Psi
_{a}\rangle $ and $|\Psi _{b}\rangle $. The first term in $\widetilde{H}_{I}$
essentially describes non-radiative transitions between exciton states due
to phonon emission. Under the assumption that the phonon modes are initially
unoccupied, we can choose the states $|\Psi _{a}\rangle $ and $|\Psi
_{b}\rangle $ such that they have a negligible propensity for nonradiative
transitions, {\it i.e.}, they are protected against single phonon emission
(recall condition 3. for ``good'' qubits in Section~\ref{qubits}).\ This
means that we can effectively set all $\gamma _{jik}$ to zero, thus
eliminating the first term in $\widetilde{H}_{I}$. This important
simplification is 
treated in detail in Section \ref{decoherence} below. Thus we are left with: 
\[
\widetilde{H}_{I}=\hbar g_{{\bf k}}^{ab}|\Psi _{a}\rangle \langle \Psi
_{b}|e^{i\phi_{t}-it(\nu _{{\bf k}}-\omega _{ab}^{e+})}\bigotimes_{k}\sum_{n,m}F_{nmk}^{ab}|n_{ak}\rangle \langle
m_{bk}|e^{i\omega _{k}^{d}(n-m)t}+{\rm H.c.} 
\]
We then tune our laser on resonance such that $\nu _{k}=\omega _{ab}^{e+}$,
and make sure that the laser spectral width is much smaller than the lowest
quantum dot phonon frequency, $\omega _{1}^{d}$. This allows us to make
the rotating wave approximation (RWA), {\it i.e.}, eliminate all terms which
rotate faster than $\omega _{1}^{d}$, which leads to

\begin{equation}
\widetilde{H}_{I}=\hbar g_{{\bf k}}^{ab}|\Psi _{a}\rangle \langle \Psi
_{b}|e^{i\phi _{t}}\bigotimes_{k}\sum_{n}F_{nnk}^{ab}|n_{ak}\rangle \langle
n_{bk}|+{\rm H.c}.  \label{eq:RWAdip}
\end{equation}
This RWA interaction Hamiltonian, Eq.~(\ref{eq:RWAdip}), is very similar to
the familiar two-level system Hamiltonian used extensively in atomic optics 
\cite{Scully:97}. However, the strength of the interaction is modulated here
by the Franck-Condon factors, $F_{nnk}^{ab}$. To allow the simplification of
the Hamiltonian from Eq.~(\ref{eq:dip}) to Eq.~(\ref{eq:RWAdip}) requires
a judicious choices of laser intensities and states. In our scheme the
occupation $n$ of all phonon modes will be initially zero. Using Eq.~(\ref
{eq:g}), it is useful to then introduce the factor

\begin{equation}
\Omega _{{\bf k}}^{ab}=g_{{\bf k}}^{ab}\prod_{k}F_{00k}^{ab}=\sqrt{\frac{
2\pi \alpha I_{{\bf k}}}{\hbar }}\prod_{k}F_{00k}^{ab}\langle \Psi
_{a}|\epsilon _{{\bf k}}\cdot ({\bf r}_{e}-{\bf r}_{h})|\Psi _{b}\rangle ,
\label{eq:dipole}
\end{equation}
which corresponds to the Rabi frequency for an on-resonant transition. $I_{
{\bf k}}$ is the laser intensity and $\alpha =e^{2}/\left( 4\pi \varepsilon
_{0}\hbar c\right) $ is the fine structure constant.

\subsubsection{Raman Transitions}

\begin{figure}[h]
\epsfig{figure=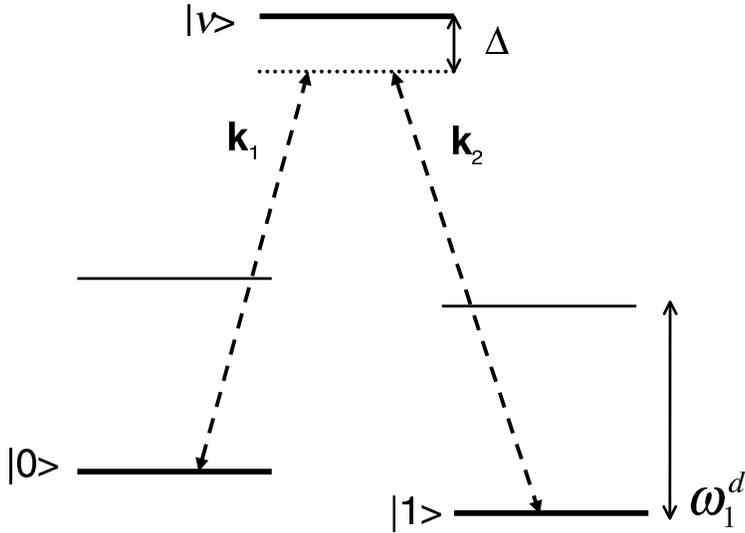,height=200pt}
\caption{ Energy level scheme for a quantum dot showing the laser fields
and transitions necessary for one-qubit operations. (Energy level spacings
are not to scale.) Levels $|0\rangle $ and $|1\rangle $ constitute the
qubit. The auxiliary level $|2\rangle $ is not involved in these transitions
and is not shown. The linear support modes are not involved either. Two
antinodal lasers, ${\bf k}_{1}$ and ${\bf k}_{2}$, allow us to perform a
Raman transition via a virtual state $|v\rangle $. Transitions occur without
changing internal phonon number, since the lasers frequency widths are
smaller than the internal phonon frequency, $\protect\omega_{1}^{d}$.}
\label{raman}
\end{figure}

Since we wish to use near degenerate states of equal parity for our qubits,
we cannot employ dipole transitions. Hence we use Raman transitions. These
connect states of equal parity via a virtual transition to a state with
opposite parity. Recall that parity is determined by $(-1)^{F_{z}}$. Suppose
we start in the $|1\rangle =|\psi _{1/2,-1/2}^{S}({\bf r}_{e})\psi
_{3/2,-3/2}^{S}({\bf r}_{h})\rangle $ state, for which $F_{z}=-2$. We can
then make transitions through a virtual level $|v\rangle $ that has opposite
parity (e.g., $F_{z}=\pm 1$), to the state $|0\rangle $ having $F_{z}=0$.
Figure~\ref{raman}a provides a schematic of the coupled QD-laser field
system, showing the levels $|0\rangle $, $|1\rangle $ and $|v\rangle $
together with the fields required to cause a Raman transition. Under the
assumptions that only two laser field modes ${\bf k}_{1}$ and ${\bf k}_{2}$
are applied, and in the rotating-wave approximation, the standard theory of
Raman transitions \cite{TLS:1976} leads to the following expression for the
Raman-Rabi frequency between an initial state $|i\rangle $ and a final state 
$|f\rangle ${\tt : 
\begin{equation}
\Omega _{{\rm Raman}}^{fi{\bf k}_{2}{\bf k}_{1}}=\frac{|\Omega _{{\bf k}
_{2}}^{fj}\Omega _{{\bf k}_{1}}^{ji}|}{\Delta }.  \label{eq:Omega-Raman}
\end{equation}
}Here$\ k_{1}$ and $k_{2}$ are chosen such that the detuning, $\Delta
=\omega _{j}-\omega _{i}-\nu _{{\bf k}_{1}}=$ $\omega _{j}-\omega _{f}-\nu _{
{\bf k}_{2}}$, $j$ is the index of an intermediate exciton state chosen to
provide a minimum value of $\Delta $. For single qubit transitions, both
lasers are aligned such that the QD is positioned at antinodes.

For QDs possessing cubic crystal structure and composed of direct band gap
materials, we have found it advantageous to use the $F_{z}=1$ and $
F_{z}=-1 $ states of the $1S_{1/2}1P_{5/2}$ multiplet as the intermediate
state. An exciton wavefunction of the $1S_{1/2}1P_{5/2}$ multiplet $\Psi
_{F_{z}}^{v}(r_{e},r_{h})$ can be expanded in terms of products of
single-particle wavefunctions $\psi _{3/2,m_{s}}^{S}(r_{e})$ and $\psi
_{5/2,m_{K}}^{P}(r_{h})$.\cite{Xia:89,Efros:96} The intermediate $F_{z}=\pm
1 $ virtual states can be written:

\[
|\Psi _{\pm 1}^{v}({\bf r}_{e},{\bf r}_{h})\rangle =-\frac{1}{\sqrt{3}}\left[
|\psi _{1/2,\mp 1/2}^{S}({\bf r}_{e})\psi _{5/2,\pm 3/2}^{P}({\bf r}
_{h})\rangle +\sqrt{2}|\psi _{1/2,\pm 1/2}^{S}({\bf r}_{e})\psi _{5/2,\pm
1/2}^{P}({\bf r}_{h})\rangle \right] 
\]
The Raman-Rabi Frequency, $\Omega _{{\rm Raman}}$, can then be adjusted by
increasing the electric field intensity and by reducing the detuning from
the intermediate level. We will describe in detail in Section \ref
{Parameters} what range of values of intensity and detuning are allowed.

\subsection{Two-qubit Operations: Coupling Quantum Dots, Quantum Supports
and Light}

\label{twoqubit}

Our two-qubit operations are equivalent to those of the Cirac-Zoller scheme
\cite{Cirac:95}. The use of optical Raman transitions to implement this
scheme has been extensively explored.\cite{Wineland:98}\ In our case, we
apply the Hamiltonian of Eq.~(\ref{eq:Hint}) with two lasers $k_{1}$ and $
k_{2}$, of frequency $\nu _{1}$ and $\nu _{2}$ respectively. For
two-qubit operations, the quantum dot is centered at an antinode of $k_{1}$
and at a node of $k_{2}$. Switching to the interaction picture and
calculating second-order transition probabilities to first order in $\eta ,$
one obtains the following effective Hamiltonian:

\begin{equation}
H_{{\rm eff}}^{nfi}=-\hbar \sum_{m}\eta _{mn{\bf k}_{2}}(a_{m}^{\dagger
}e^{-i\omega _{m}t}+a_{m}e^{i\omega _{m}t})\Omega _{{\rm Raman}}^{fi{\bf k}
_{2}{\bf k}_{1}}|\Psi _{f}\rangle _{n}\langle \Psi _{i}|e^{i(\omega
_{f}-\omega _{i})t}e^{-i(\left( \nu _{1}-\nu _{2}\right) t+\phi _{2}-\phi
_{1})}+{\rm H.c.}  \label{eq:Hnfi-eff}
\end{equation}
Note that the nodal and antinodal lasers result in an effective Hamiltonian
in which $\eta $ depends only on the nodal laser $k_{2}$. This differs from
the effective Hamiltonian derived for Raman transitions when travelling
waves are used \cite{Wineland:98}. The lasers are chosen to have a net red
detuning, $\omega _{f}{}-\omega _{i}-\upsilon _{1}+\upsilon _{2}=-\omega
_{m^{\prime }}$.\ In the RWA (i.e., eliminating all terms rotating at $
2\omega _{m}$), with $\phi _{2}-\phi _{1}=\pi ,$ this yields

\begin{equation}
H_{{\rm eff}}^{nfi}=\Omega _{nm^{\prime }}^{fi{\bf k}_{2}{\bf k}_{1}}|\Psi
_{f}\rangle _{n}\langle \Psi _{i}|a_{m^{\prime }}+{\rm H.c.}
\end{equation}
where

\begin{equation}
\Omega _{nm^{\prime }}^{fi{\bf k}_{2}{\bf k}_{1}}=\eta _{mn{\bf k}
_{2}}\Omega _{{\rm Raman}}^{fi{\bf k}_{2}{\bf k}_{1}}.  \label{eq:omega}
\end{equation}
This combined QD-linear support operation transfers the $n^{{\rm th}}$ QD
from state $i$ to $f$, with an accompanying change of one quantum in the
phonon mode $m^{\prime }$ of the support. A schematic representation of this
operation for the qubit states $|\psi _{i}\rangle =|0\rangle $ and $|\psi
_{f}\rangle =|1\rangle $ is shown in Figure~\ref{levels}. Choosing
interaction times such that $t=k\pi /(2\Omega _{nm^{\prime }}^{fi{\bf k}_{2}
{\bf k}_{1}})$ where $k$ is an integer specifying the pulse duration, we can
write the unitary operator $\exp \left( -\frac{i}{\hbar }H_{{\rm eff}
}^{nfi}t\right) =U^{nfi}(t)$ as 
\begin{equation}
U_{k}^{nfi}=\exp \left[ -i\frac{k\pi }{2}(|\Psi _{f}\rangle _{n}\langle \Psi
_{i}|a+{\rm H.c.})\right]   \label{eq:U_nfi}
\end{equation}
In order for the Cirac-Zoller scheme to be successful, the phonon mode of
interest, $m$, must start with zero occupation. The applied operations take
advantage of the fact that the zero occupation phonon state is annihilated
by the lowering operator: $a|0\rangle =0$. The sequence of unitary
operations $U_{{\rm C-phase}}\equiv U_{1}^{n10}U_{2}^{n^{\prime
}20}U_{1}^{n10}$ then results in a controlled-phase operation between
quantum dots $n$ and $n^{\prime }$, i.e., it causes the second qubit $
n^{\prime }$ to gain a phase of $-1$ if the first qubit $n$ is in the $
|1\rangle $ state, and no additional phase if the first qubit is in $
|0\rangle $. This is equivalent to the matrix operator $I-2|1\rangle
_{n^{\prime }}|1\rangle _{n}\ _{n^{\prime }}\langle 1|_{n}\langle 1|$. The
time required to perform $U_{{\rm C-phase}}$ is then $2\times \left( \frac{
\pi }{2}\right) $ for ion $n$ plus $1\times \pi $ for ion $n^{\prime }$,
i.e.,

\begin{equation}
\tau _{{\rm C-phase}}\equiv (\Omega _{2}^{nn^{\prime }})^{-1}=\pi (\Omega
_{nm^{\prime }}^{fi{\bf k}_{2}{\bf k}_{1}})^{-1}+\pi (\Omega _{n^{\prime
}m^{\prime }}^{fi{\bf k}_{2}{\bf k}_{1}})^{-1}  \label{eq:taucph}
\end{equation}
Since the ions $n$ and $n^{\prime }$ are identical we will use the
approximation that $\Omega _{nm^{\prime }}^{fi{\bf k}_{2}{\bf k}_{1}}\approx
\Omega _{n^{\prime }m^{\prime }}^{fi{\bf k}_{2}{\bf k}_{1}}$ in the
remainder of this work, hence

\begin{equation}
\Omega _{2}^{nn^{\prime }}\approx \frac{1}{2\pi }\Omega _{nm^{\prime }}^{fi
{\bf k}_{2}{\bf k}_{1}}=\frac{1}{2\pi }\Omega _{2}.  \label{eq:approx}
\end{equation}
The inverse of the average rate $\Omega _{2}^{nn^{\prime }}$ can then be
taken as a measure of the gate time, i.e., of the time for the two-qubit
controlled phase operation. We define $\Omega _{2}$ as the sideband
interaction strength,

\begin{equation}
\Omega _{2}\simeq \eta _{mn{\bf k}_{2}}\Omega _{{\rm Raman}}^{fi{\bf k}_{2}
{\bf k}_{1}}.  \label{eq:omega2}
\end{equation}
Calculation of $\Omega _{{\rm Raman}}^{fi{\bf k}_{2}{\bf k}_{1}}$ was
described above in Section~\ref{single} [Eq.~(\ref{eq:Omega-Raman})]. We can
obtain the Lamb-Dicke parameter $\eta _{m^{\prime }n{\bf k}_{2}}$ from the
decomposition in Eq.~(\ref{eq:LD2}). This now allows specific evaluation of
the contribution from the linear support to the Lamb-Dicke parameter $\eta $. As described in Section \ref{approx}, we approximate $\eta $ as being
independent of the specific dot and from now on will drop the dot index $n$.

\subsection{Input and Output}

Since the qubit states do not include the ground state of the quantum dot,
initialization will generally require a transformation from the ground state
of no exciton to the defined qubit state. This can be accomplished by
applying magnetic fields which will mix dark and light states allowing for
optical transitions. If the magnetic field is then adiabatically removed,
one is left with population in the dark exciton state only. Qubit
measurements can be made by using a cycling transition, in analogy to ion
traps \cite{Wineland:98}.

\begin{figure}[h]
\epsfig{figure=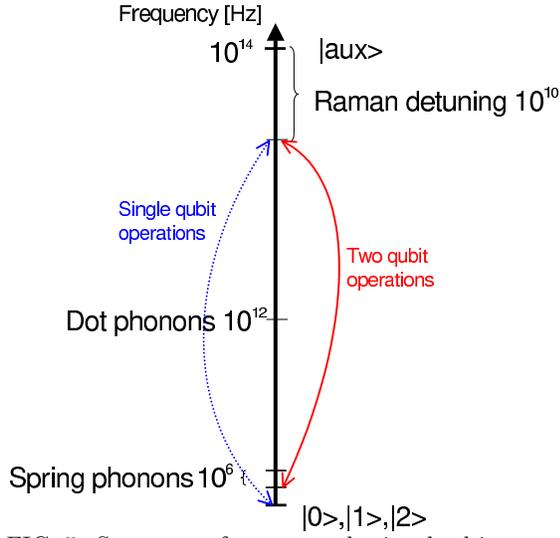,height=200pt}
\caption{Summary of energy scales involved in our proposal for single- and
two-qubit operations. The states $|0\rangle$,$|1\rangle$, and
$|2\rangle$ appear degenerate on the scale of this figure.}
\label{energies}
\end{figure}

We conclude this section by summarizing in Fig.~\ref{energies} the relative
energy scales involved in our proposal.

\section{Feasibility of Quantum Logic}
\label{feasibility}

In this section we address in detail the question of the limitations
imposed on our system by various physical constraints. We start by
considering the issue of decoherence due to coupling of excitons to
the internal nanocrystal phonon modes, and propose a solution to this
problem. We then study the issues of scaling arising from the
trade-off between massiveness of the support, laser intensity, and the
need to maintain a large ratio of operations to exciton recombination
time. We find that the allowed size of our proposed quantum computer
depends on the assumed threshold for fault-tolerant computation.

\subsection{Decoherence}

\label{decoherence}

According to current analysis of experiments on nanocrystal quantum
dots \cite{Takagahara:96}, exciton dephasing derives predominantly from the
diagonal phonon exciton coupling term of Eq.~(\ref{eq:couple}): 
\begin{equation}
\sum_{j,k}|j\rangle \langle j|(\gamma _{kj}^{\ast }b_{k}^{\dagger }+\gamma
_{kj}b_{k}).  \label{eq:H_I}
\end{equation}
Here $\gamma _{kj}$ is the self-coupling of an exciton state $j$ via phonon $
k$, and $b_{k}$ is the lowering operator for the $k^{{\rm th}}$ phonon mode
in the ground electronic state. In the ground state the coupling is zero,
and all excited electronic states have potential energy surfaces which are
shifted with respect to this ground state. We desire to eliminate
dephasing due to the first-order phonon exciton interaction. In the typical
experimental situations in which dephasing has been studied in the past,
dephasing occurs on a timescale of nanoseconds for small dots \cite
{Takagahara:96}. This rate is extremely rapid compared to the
experimental recombination time of dark states ($\sim 10^{-6}$ s for direct
band gap materials such as CdSe\cite{Efros:96}). The reason for such fast
dephasing is twofold. First, the vibrational stationary state of the first
electronic level becomes a moving vibrational wave packet on the upper
electronic surface, because the spectral width of the pulse is too broad to
distinguish vibrational eigenstates. Second, the QD is embedded in a solid
state medium where the vibrations of the nanocrystal are then coupled to
vibrations of the larger lattice. The phonons of the QD can be treated as
analogous to damped cavity modes in atomic optics\cite{Scully:97}. In the
case of strong coupling, one finds from numerical simulation that the
dephasing between any two states $j$ and $j^{\prime }$ is related to the
rate of phonon mode excitation. The latter is proportional to $|\gamma
_{kj}-\gamma _{kj^{\prime }}|^{2}$ for each mode $k$ \cite{Takagahara:96}.
This conclusion of fast dephasing agrees with the analogous result for a
leaky optical cavity\cite{Scully:97} as well as with the results of
experimental \cite{Banin:97} and theoretical \cite{Takagahara:96} analysis
for embedded semiconductor nanocrystals.

The dephasing can be reduced in three ways. First, the coupling of the QD
phonon modes to external phonon and photon modes can be reduced by judicious
choice of nanocrystal geometry and material. In our case the QD can
dissipate phonon modes only to the support. In the limit of no coupling to
external modes, there will be no dephasing but the time required for
recurrences could limit our gate repetition rate. Although the oscillations
will be fast, the oscillations for different phonon energies will be
incommensurate with one another. This could introduce a slow quantum
beating between ground and excited electronic states, which would have the
undesirable consequence of requiring gate durations to equal a full beat
cycle. 

Second, one can find a set of electronic states $|j\rangle $ such that

\begin{equation}
\gamma _{kj}-\gamma _{kj^{\prime }}=0\text{ \ \ }\forall j,j^{\prime },
\text{and }\forall k.
\end{equation}
Physically this condition represents a set of electronic states which create
the same potential energy surface for nuclear motion. This elimination of
decoherence by degeneracy is an example of a decoherence-free subspace \cite
{Duan:97PRL,Zanardi:97c,Lidar:PRL98}. The Jahn-Teller effect implies
that no two such states 
should exist, because there will always be a phonon mode which will
distinguish between these states due to non-linearity\cite{Bunker}. However,
in the linear approximation we have

\begin{equation}
\gamma _{kj}=\langle j|\gamma _{k}({\bf r}_{h})+\gamma _{k}({\bf r}
_{e})|j\rangle .  \label{eq:gamma_el}
\end{equation}
The deformation potential coupling operator, $\gamma _{k}(r)$, is a function
of the phonon modes and is expressed as 
\begin{equation}
\gamma _{k}({\bf r}) \equiv \gamma _{nlm}({\bf r})=E_{d}\nabla \cdot
{\bf u}_{nlm} {\bf (r)},
\end{equation}
where $E_{d}$ is the deformation potential, and $u_{nlm}(r)$ is the
coordinate representation of the normalized spheroidal phonon mode of level $
n$ with angular momentum $l$ and projection $m$. Following Takagahara \cite
{Takagahara:96}, the spheroidal modes can be written as

\begin{equation}
{\bf u}_{nlm}{\bf (r)=}\sqrt{\frac{\hbar }{2\rho \omega _{nlm}}}(p_{nl}{\bf L
}_{lm}(h_{nl}{\bf r})+q_{nl}{\bf N}_{lm}(k_{nl}{\bf r})),  \label{eq:u_nlm}
\end{equation}
where $\rho $ is the nanocrystal density, $\omega _{nlm}$ is the frequency
of the spherical phonon $nlm$, $L_{lm}(hr)=\frac{1}{h}\nabla \Psi _{lm}(hr)$, $N_{lm}(kr)=\frac{1}{k}\nabla \times \nabla \times r\Psi _{lm}(kr)$, and $
\Psi (kr)=j_{l}(kr)Y_{l}^{m}(\Omega )$. $j_{l}\left( r\right) $ is an $l^{
{\rm th}}$ order spherical Bessel function, $Y_{l}^{m}(\Omega )$ is a
spherical harmonic, $k_{n}$ and $h_{n}$ satisfy stress-free boundary
conditions at the surface, and $p_{nl},q_{nl}$ are determined by
normalization. One can then write 
\begin{equation}
\gamma _{nlm}({\bf r})=-E_{d}\sqrt{\frac{\hbar }{2\rho \omega _{nlm}}}
p_{nl}h_{nl}j_{l}(h_{nl}r)Y_{l}^{m}(\Omega ).  \label{eq:gamma_nlm}
\end{equation}
For a cubic, direct gap nanocrystal, the states $|j\rangle $
are states of well defined angular momentum projection. Since the
$Y_{l}^{m}$ in Eq.~(\ref{eq:gamma_nlm}) connects states with equal projection, the
only 
phonon modes which can have non-zero matrix elements in Eq.~(\ref{eq:gamma_el})
are those with $m=0$ \cite{Rose:book}. The resulting matrix elements will be independent of
the sign of the exciton angular momentum projection $F_{z}$ [Eq.~(\ref{eq:F})], i.e., the $m=0$ phonon modes cannot distinguish between exciton states
having $F_{z}$ or $-F_{z}$. Therefore, in the linear approximation, the
states$|\Psi _{-2}(r_{e},r_{h})\rangle $ and $|\Psi _{2}(r_{e},r_{h})\rangle 
$ will not dephase with respect to each other. Recall that we took these
states as our qubit state $|1\rangle $ and auxilary state $|2\rangle $
states [Eqs.~(\ref{eq:|1>}),(\ref{eq:|2>})].

Third, and most importantly, one can change the way in which transitions are
made. In the above two situations, the motional wave packet of one electronic
surface is transferred to another electronic surface without changing shape,
i.e., the Franck-Condon approximation holds. However, such a transition
requires either a broad laser or a fast excitation. This is not actually
the regime of relevance here. The scheme outlined in this work requires
selective excitations of sidebands whose energy separation is orders of
magnitude smaller than the quantum dot phonon energies (see Fig.~\ref
{energies}). Therefore, we will be performing transitions from one
vibrational eigenstate to another vibrational eigenstate. Such transitions
were described in Section \ref{operations} with respect to the ground vibrational state. Consequently, the scheme proposed in this work is not
affected by fast phonon dephasing.

Exciton states recombine and thus decay to the ground state by both
spontaneous emission of photons and phonons. We denote the recombination
lifetime $\tau _{{\rm re}}$. Nanocrystals are known to have dark state
recombination times ranging from $10^{-6}$s to $10^{-3}$s, depending on
the material chosen \cite{Efros:96,Leung:98,Heckler:99,KLeung:97}. In our
system, $\tau _{{\rm re}}$ will be the fastest decoherence time for
individual qubits. One could potentially suppress radiative recombination by
placing the whole system in a cavity\cite{cavity}. \ Classical
calculations of Roukes and co-workers show that nanoscale rods at low
temperature have high $Q$ values: $Q\geq 10^{10}$ \cite{Roukes}. This
implies that the rods are only very weakly coupled to their environments and
we can therefore assume that in the quantum regime, the high $Q$ will lead
to favorably long decoherence times. Another possible source of decoherence
is laser scattering from the support (as opposed to the QDs). The magnitude
of scattering is dependent on the difference between the spectra of the
quantum dots and the electronic and vibrational modes of the support. A
detailed analysis of this potentially important decoherence mechanism is
beyond the scope of this paper, due to the many possible materials available
for both linear supports and quantum dots. Ideally, one would like to choose
a support which has an optical window at the frequencies of the lasers used
to perform the qubit operations. Estimations based on the Raman transitions
proposed above suggest that this optical window needs to be between 0.1 - 1
eV and possess a minimal width of 0.1 meV. Conversely, one can take
advantage of the optical tunability of the QDs via their size to construct a
QD of such size that its transitions are compatible with a specific optical
window suggested by the material properties of the support.

\subsection{Parameter Space}

\label{Parameters}

We now explore for what range of physical parameters quantum computation is
possible within our proposed scheme, by estimating the two-qubit gate
fidelity, ${\cal F}$. This fidelity is defined
as the trace overlap between the desired final and the achieved final state: 
${\cal F}=\min_{\rho _{0}}{\rm Tr}A\rho _{0}A^{\dagger }B(\rho _{0})$, where $A$
is the exact unitary operator for the gate, $B$ is a superoperator
describing the actual evolution of the system which takes the initial density
matrix $\rho _{0}$ to a final density matrix $\rho _{f}$, and $\rho _{0}$
ranges over all possible input states. For our two-qubit operations
described in Section~\ref{twoqubit}, the fundamental operation is the
population transfer to the red side band,
$A=U_{1}^{n10}$[Eq.~(\ref{eq:U_nfi})]. $B$ describes both the unitary evolution caused by application of the
lasers and the decoherence due to loss of quantum information to the environment. Note
that even without decoherence and unknown laser noise, ${\cal F}$ can still
be less than unity, due to deviations from the approximations used to derive 
$A$. Most importantly, deviations from the rotating wave approximation can
lead to unwanted population in spectator states.

\begin{figure}[h]
\epsfig{figure=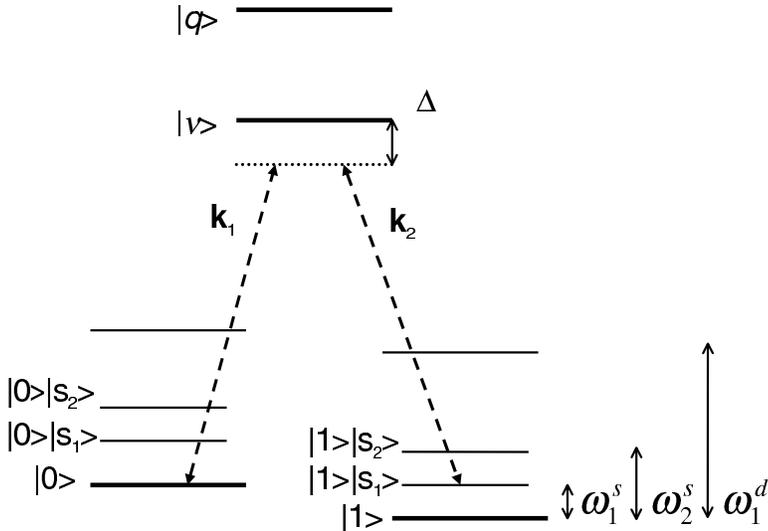,height=200pt}
\caption{
Energy level scheme for a quantum dot on the linear support showing laser
fields necessary for implementation of two-qubit operations. As described in
the text, the use of nodal and antinodal lasers allows us to selectively
transfer population from $|0\rangle $ to the lowest energy phonon sideband
of $|1\rangle $ (labelled $|1\rangle |s_{1}\rangle $) via a Raman
transition, without transferring population to the carrier. The minimum
phonon frequency is denoted $\protect\omega _{1}^{s}$. Non-resonant
transitions to higher energy phonon sidebands, ($|1\rangle |s_{2}\rangle $)
constitute the main source of error in the proposed gates. For very high
laser intensities, non-resonant quadrupolar transitions to higher level
states (represented by the state $|q\rangle $) also become important.}
\label{levels}
\end{figure}

The resulting value of ${\cal F}$ is determined by two constraints: the
decoherence time of the system and the spectral resolution of the gate. As a
result of the use of the phonon bus in the two qubit gate construction, both
this proposal and the ion trap proposals \cite{Cirac:95,Sorensen:99} have
gate times that are dependent on the number of qubits. Therefore the quantity
of interest is the fidelity of a sideband operation on an array of $N$
quantum dots. First, we note that ${\cal F}$ is limited by the
recombination time of the qubit states, $\tau _{{\rm re}}=1/\Gamma $.
Naturally, the recombination time $\tau _{{\rm re}}$ must be larger than
the sideband operation time, $\tau _{A}=\pi /(2\Omega _{2})$, if the
operation is to be successful. One can then define an upper limit on the
fidelity that takes into account the statistically independent
recombination of the exciton 
states of all $N$ quantum dots. This background fidelity assumes
that $\tau _{{\rm re}}$ is the same for all $N$ quantum dots and it does
not account for errors deriving from the interactions with the driving laser
field. The background fidelity is then

\begin{equation}
{\cal F}{\cal \approx }1-\frac{\tau _{A}}{\tau _{{\rm re}}}=1-\frac{\pi
N\Gamma }{2\Omega _{2}}.  \label{eq:bfid}
\end{equation}
${\cal F}$ is also limited by the spectral resolution. We assume that the
difference frequency between the lasers is tuned to be resonant with the
energy difference between the two states of interest, Figure~\ref{levels}.
The energy difference between the states includes the relative Stokes shift
induced by both the lasers and the internal phonons. Omitting the adjustment
of the laser frequency for the Stokes shifts will lead to gates of reduced
fidelity \cite{Blatt:2000,Knight:2000}. The primary concern is that the phonon modes be spectrally resolvable. As
described in Section~\ref{overview}, the use of nodal and antinodal lasers
allows us to transfer population to the vibrational sideband and at the same
time forbid population transfer to the carrier, i.e., to the excitonic
states of the QD. This constrains the operation frequency $\Omega _{2}$ to
be smaller than the separation between phonon modes, $\Delta \omega =\omega
_{m+1}^{s}-\omega _{m}^{s}$. For small $m$, and for $\Delta \omega \approx
\omega _{1}^{s}$, we can then write this constraint as 
\begin{equation}
\Omega _{2}<\omega _{1}^{s}.  \label{sr}
\end{equation}
The population transfer to the off-resonant state will be of the order of $\frac{
4g^{2}}{4g^{2}+(\delta )^{2}}$, the result for a two level
system interacting with a periodic perturbation of strength $g$ that is
off-resonant by a frequency difference $\delta $ \cite{Scully:97}. In
our case, $g=\Omega _{2} 
$ and $\delta =\Delta \omega \approx \omega _{1}^{s}$. However, numerical
calculations \cite{Blatt:2000} have shown that the population transfer to
the off-resonant state for an ion trap system is more accurately estimated
by $2\frac{g^{2}}{4g^{2}+(\delta )^{2}}.$ Note that in Ref. \cite{Blatt:2000}
g=$\Omega /2,$ the coupling to the carrier transition, and $\delta =\omega
_{z}$, the ion trap mode vibrational frequency. The advantage of the standing
wave laser configuration is now apparent. For the travelling wave laser
configuration, the population transfer to the off-resonant state is $2\frac{
(\Omega _{2}/\eta )^{2}}{4(\Omega _{2}/\eta )^{2}+(\omega _{1}^{s})^{2}}
\approx 2\left( \frac{\Omega _{2}}{\eta \omega _{1}^{s}}\right) ^{2}$ (or,
in the notation of Ref. \cite{Blatt:2000}, $\frac{1}{2}\left( \frac{\Omega }{
\omega _{z}}\right) ^{2}).$ However, in the standing wave configuration,
one finds that the population transfer is $2\frac{(\Omega _{2})^{2}}{
4(\Omega _{2})^{2}+(\omega _{1}^{s})^{2}}$ $\approx 2\left( \frac{\Omega _{2}
}{\omega _{1}^{s}}\right) ^{2}$. Since $\eta \ll 1$ the off-resonant
population transfer is significantly reduced when one uses the standing wave
configuration. We can now write down a fidelity which takes into account
both the background fidelity, Eq.~(\ref{eq:bfid}), and the population loss
to the most significant spectator state. The fidelity per sideband
operation $A$ is then

\begin{equation}
{\cal F} \approx  1-\frac{\pi
N\Gamma }{2\Omega _{2}}-2\left( \frac{\Omega _{2}}{\omega _{1}^{s}}\right)
^{2}  \label{eq:fidelity}
\end{equation}
in the standing wave configuration. We emphasize again that if the laser
fields are used in a traveling wave configuration, the fidelity is
significantly decreased due to transitions to the carrier state, resulting
in ${\cal F}\simeq $ $1-\frac{\pi N\Gamma }{2\Omega _{2}}-4(\frac{\Omega _{2}
}{\eta \omega _{1}^{s}})^{2}$.

One can now maximize the fidelity with respect to the coupling strength $
\Omega _{2}$ for a sideband operation made on an $N$ qubit array. Since

{\tt 
\begin{equation}
\Omega _{2}^{{\cal F}\max }=\left( \frac{\pi (\omega _{1}^{s})^{2}N\Gamma }{8
}\right) ^{1/3},  \label{maximum}
\end{equation}
}
for given $\Gamma $ and $\omega _{1}^{s}$, the maximum
fidelity can be written as

\begin{equation}
{\cal F}_{\max }{\cal =}1-3\left( \frac{\pi N\Gamma }{2\sqrt{2}\omega
_{1}^{s}}\right) ^{2/3}.  \label{fmax}
\end{equation}
Evaluation of the optimal operation frequency depends then only on the
factors in Eq.~(\ref{maximum}). Inspection of the contributions to $\Omega
_{2}$ [Eq.~(\ref{eq:omega2})] shows that the underlying adjustable
parameters controlling the fidelity in general, Eq.~(\ref{eq:fidelity}), are
the intensities $I_{1}$, $I_{2}$, the detuning $\Delta $, and the Lamb-Dicke
parameter $\eta $. However, the schematic shown in Figure~\ref{levels} shows
that there are some additional constraints. Thus, it is essential that the
inequality $\omega _{1}^{s}<\Delta <\omega _{1}^{d}$ be satisfied in order
to avoid unwanted coupling to both the internal phonons and the linear
support phonons. In addition, we require that both $|\Omega _{{\bf k}
_{1}}^{ji}|$ and $\eta _{01{\bf k}_{2}}|\Omega_{{\bf k}_{2}}^{fj}|$ are
smaller than $\Delta $, in order to avoid populating the intermediate level, 
$|\Psi _{j}\rangle $. Hence, the internal phonon energies define an energy
scale which also constrains our system (see Fig.~\ref{energies}).

Furthermore, combining Eqs.~(\ref{eq:Omega-Raman}),~(\ref{eq:omega2}), and~(\ref{maximum}), one finds that 
\begin{equation}
|\Omega _{{\bf k}_{1}}^{ji} \eta _{01{\bf k}_{2}}\Omega _{{\bf k}
_{2}}^{fj}|=\Delta \left( \frac{\pi (\omega _{1}^{s})^{2}N\Gamma }{8}\right)
^{1/3}.  \label{eq:omegacond}
\end{equation}
Analysis of three level systems has shown that in order to maximize Raman
population transfer between states $|\Psi _{i}\rangle $ and $|\Psi
_{f}\rangle $, the coupling strength between $|\Psi _{i}\rangle $ and $|\Psi
_{j}\rangle $ and $|\Psi _{f}\rangle $ and $|\Psi _{j}\rangle $ should be
equal \cite{TLS:1976}. In the atomic case this usually implies that the
respective Rabi frequencies between electronic states, $|\Omega _{{\bf k}
_{1}}^{ji}|$ and $|\Omega _{{\bf k}_{2}}^{fj}|$, are equal. However, in our
case, with the use of nodal and anti-nodal lasers and coupling to the
support phonons, the equivalent condition is that 
\begin{equation}
|\Omega _{{\bf k}_{1}}^{ji}|=\eta _{01{\bf k}_{2}}|\Omega _{{\bf k}
_{2}}^{fj}|.  \label{eq:equality}
\end{equation}
Therefore, manipulation of Eqs.~(\ref{eq:dipole}), (\ref{eq:omegacond}) and
(\ref{eq:equality}), allows us to determine the intensity values, $I_{1}$
and $I_{2}$, necessary for maximum fidelity operations:

\begin{eqnarray}
I_{1} &=&\frac{\hbar \Delta }{2\pi \alpha }\left( \frac{\pi (\omega
_{1}^{s})^{2}N\Gamma }{8}\right) ^{1/3}\frac{1}{|\langle \Psi _{j}|\epsilon
_{1}\cdot ({\bf r}_{e}-{\bf r}_{h})|\Psi _{i}\rangle
\prod_{l}F_{00l}^{ji}|^{2}}  \label{eq:intense} \\
I_{2} &=&\frac{\hbar \Delta }{2\pi \alpha }\left( \frac{\pi (\omega
_{1}^{s})^{2}N\Gamma }{8}\right) ^{1/3}\frac{1}{\eta _{01{\bf k}_{2}}^{2}}
\frac{1}{|\langle \Psi _{f}|\epsilon _{2}\cdot ({\bf r}_{e}-{\bf r}
_{h})|\Psi _{j}\rangle \prod_{l}F_{00l}^{fj}|^{2}}  \label{eq:intensity_max}
\\
&=&\frac{\Delta \lambda L}{2\pi \alpha }\left( \frac{\pi (\omega
_{1}^{s})^{5}N\Gamma }{8}\right) ^{1/3}\frac{1}{|{\bf k}_{2}{\bf |}^{2}}
\frac{1}{|\langle \Psi _{f}|\epsilon _{2}\cdot ({\bf r}_{e}-{\bf r}
_{h})|\Psi _{j}\rangle \prod_{l}F_{00l}^{fj}|^{2}}  \nonumber
\end{eqnarray}
Equations (\ref{fmax}) , (\ref{eq:intense}), and (\ref{eq:intensity_max})
summarize the limits to implementation of this quantum dot-quantum linear
support scheme. To maximize the fidelity we need to increase the frequency
of the phonon bus, $\omega _{1}^{s}$. However, as this frequency
increases, the increased intensity necessary to reach the maximum fidelity
will lead to unwanted evolutions not considered in our simple fidelity
equation, Eq.~(\ref{fmax}). These unwanted evolutions include quadrupolar
excitation to higher electronic states. Such transitions will not be
removed by the use of nodal and anti-nodal lasers \cite{James:98}.
Consequently, it is useful to define maximal laser intensities, $I_{1}^{\max
}$ and $I_{2}^{\max },$ such that Eq.(\ref{fmax}) is valid for $
I_{1}<I_{1}^{\max }$ and $I_{2}<I_{2}^{\max }$. Therefore the magnitude of $
\omega _{1}^{s}$ is restricted and thereby imposes a constraint on the
phonon spectrum of the linear support. Furthermore, scalability is also
limited by the additional unwanted evolutions, since the required intensity
to achieve maximum fidelity also increases with $N$.

There are also physical constraints on the density and length of the
support. We assume that the minimal linear density would be provided by a
chain of carbon atoms, for which we estimate $\lambda _{{\rm 0}}$=10 amu/\AA
. The length of the support, $L$, is determined by the number of QDs, $N$,
and by the spatial width of the laser, $l$. Thus for identical dots, we have 
$L=lN$. One could use QDs having no spectral overlap, obtained by making the
dots of sufficiently different sizes, in order to achieve more qubits per
unit length.

Notice that when one rewrites $L$ in terms of $N$ that the intensity of the
nodal laser, $I_{2},$ has a stronger $N$ dependence than the intensity of the
anti-nodal laser, $I_{1}$. Physically, this is due to the increased inertia
of the system and is quantified by the Lamb-Dicke parameter $\eta $. From
Eq.~(\ref{eq:LD}) and Section (\ref{support}), $\eta \propto M^{-1/2}$ 
where $M$ is the total mass of the support, $M=\lambda L,$ so that $
I_{2}\propto M$ [Eq.(\ref{eq:intensity_max})]. This coupled with our
expression for the maximal gate fidelity, Eq.(\ref{fmax}), yields $
I_{2}\propto N^{4/3}$. Note that we have used the Lamb-Dicke parameter
consistent with the definition made in Ref. \cite{Blatt:2000} which has an
inverse $N$ dependence. 

To determine the scalability of our system, we examine the maximum number of
QDs which can be sustained by a support having given values of $\omega
_{1}^{s}$ and $\lambda /\lambda _{\min },$ and provide an acceptable level
of fidelity for the sideband operation, $A=U_{1}^{n10}$. We do this by
requiring the following three constraints to be simultaneously satisfied: i) 
${\cal F}={\cal F}_{\max }$, ii) $I_{2}\leq I$ $_{2}^{\max }$, and iii) $
{\cal F}_{\max }>1-\varepsilon ,$ where $\varepsilon $ can be thought of as
the error rate per gate frequency. The first condition states that maximum
gate fidelity, Eq.~(\ref{fmax}), is achieved given the support and quantum
dot parameters $\omega _{1}^{s}$ and $\Gamma $, respectively. The second
condition states that the laser frequency $I_{2}$ should not exceed the
maximum allowed value (see above). Eq.~(\ref{eq:intensity_max}) together
with the considerations in the previous paragraph shows that $I_{2}$ is
dependent on the number of quantum dots $N$. For the range of parameters
considered here, $I_{1}$ is always smaller that $I_{1}^{\max }.$ Hence the
maximum number of qubits will be determined by the intensity threshold of
the system at a node of the laser field. The third condition ensures that
one is able to perform an operations with success greater than a certain
threshold value (equal to $1-\varepsilon $). Combining these inequalities
leads to limits on the number of qubits for a given system$.$ \ Conditions
i) and iii) can be manipulated to yield the following constraint on $N$:

\begin{equation}
N\leq \omega _{1}^{s} \left( \frac{2\varepsilon }{3}\right) ^{3/2}\frac{1}{\pi
\Gamma }.  \label{eq:smallw}
\end{equation}
On the other hand, conditions i) and ii) yield a constraint with an
inverse power dependence on $\omega _{1}^{s}$. One finds that:

\begin{equation}
N\leq (\omega _{1}^{s})^{-5/4} (I_{2}^{\max })^{3/4}\left( \frac{8}{\pi \Gamma }
\right) ^{1/4}\left( \frac{2\pi \alpha }{\Delta \lambda l}|{\bf k}_{2}{\bf |}
^{2}|\langle \Psi _{f}|\epsilon _{2}\cdot ({\bf r}_{e}-{\bf r}_{h})|\Psi
_{j}\rangle \prod_{l}F_{00l}^{fj}|^{2}\right) ^{3/4}  \label{eq:bigw}
\end{equation}
One can then analyze $N_{\max }$, the maximum allowed value of $N,$ as a
function of the linear support frequency $\omega _{1}^{s}$. The
combination of Eqs.~(\ref{eq:smallw}) and (\ref{eq:bigw}) results in
a cusped function for $N_{\max }$ and is discussed in detail in Section \ref
{estimates} for both a direct band gap semiconductor (CdTe) and an indirect
band gap semiconductor (Si).

The above discussion has focused on the fidelity for a single component
operation, $A=U_{1}^{n10},$ of the C-phase gate $U_{{\rm C-phase}}\equiv
U_{1}^{n10}U_{2}^{n^{\prime }20}U_{1}^{n10}.$ We have termed this a sideband
operation fidelity. Similar arguments may be made to derive the full
C-phase gate fidelity, resulting in the expression

\begin{equation}
{\cal F}_{{\rm C-phase}}{\cal \approx }1-\frac{2\pi N\Gamma }{\Omega _{2}}
-4\left( \frac{\Omega _{2}}{\omega _{1}^{s}}\right) ^{2}.
\end{equation}
This is lower than the sideband operation fidelity, both because of the
effect of multiple couplings to spectator states and because of an increased
operation duration ($\tau _{{\rm C-phase}}=4\tau _{A}$). In the presentation
of numerical results in the next section we shall refer only to the
prototypical sideband operation fidelity, ${\cal F}$ of
Eq.~(\ref{eq:fidelity}).

\section{Numerical Estimates for Specific Nanocrystal Systems}

\label{estimates}

\subsection{CdTe}

CdTe nanocrystals are an example of direct band gap cubic crystal
semiconductors QDs. Using parameters found in Landolt-Bornstein \cite
{LandoltB}, we have performed the calculations summarized in the previous
sections, using previous results of Efros \cite{Efros:96} and Takagahara 
\cite{Takagahara:96} for the EMA analysis. Although the dark states have
infinite lifetimes in the EMA approximation, both experimental \cite
{Kapitonov:99} and tight binding calculations for the analogous CdSe system 
\cite{Leung:98} yield radiative recombination rates $\Gamma \sim 10^{6}$.

We analyze here nanocrystals with $R=20$ \AA. For this size, the frequency
of the lowest internal phonon is $\omega _{1}^{d}=2.45\times 10^{12}$Hz.
Assuming a minimal separation of a single order of magnitude between
the energy spacings $\omega ^{d}$ and $\Delta $, we set $\Delta =10^{11}$Hz. The energy separation between the 1S$_{1/2}$1S$_{3/2}$ and 1S$
_{1/2}$1P$_{5/2}$ multiplets is $\sim 0.4$ eV in the EMA, which leads to
required wave vectors $k_{1}\simeq k_{2} \simeq 2.1 \mu$ m$^{-1}$ for
the irradiating lasers in the two-qubit gates. For the specific CdTe
states introduced above (Section \ref{qubits}), we calculate the dipole
moments to be $\langle \Psi _{-1}^{\rm aux}({\bf r}_{e},{\bf r}_{h})|\epsilon
_{2}\cdot {\bf r}|0\rangle =0.11$ $R$ , $\ \langle \Psi _{-1}^{\rm aux}({\bf r}
_{e},{\bf r}_{h})|\epsilon _{1}\cdot {\bf r}|1\rangle =-0.013$ $R$ where $
\epsilon _{2}=\frac{1}{\sqrt{2}}(\widehat{x}+i\widehat{y})$ and $\epsilon
_{1}=\frac{1}{\sqrt{2}}(\widehat{x}-i\widehat{y}).$ Furthermore we have
calculated the Frank Condon overlap to be $\prod_{l}F_{00l}^{\rm aux0}=0.98$ and 
$\prod_{l}F_{00l}^{\rm aux1}=$ 0.98. We assume that the spatial width of our
lasers is diffraction limited. A reasonable estimate of this width is then $
l $ = 3 $\mu $m. At constant frequency, an increase in the number of
qubits requires an increase in the laser intensity in order to maintain
maximum fidelity operations.\ We estimate that $I_{2}^{\max }=$ 10$^{12}$W/cm
$^{2}$ is the intensity at which the nonresonant quadrupole interactions
begin to rise in CdTe quantum dots. However, the intensity could
potentially have stricter limitations depending on the spectra of the
specific support chosen. As mentioned above for the range of parameters we
have examined, the intensity of the anti-nodal laser is weak enough that it
does not lead to unwanted time evolutions.

\begin{figure}[h]
\epsfig{figure=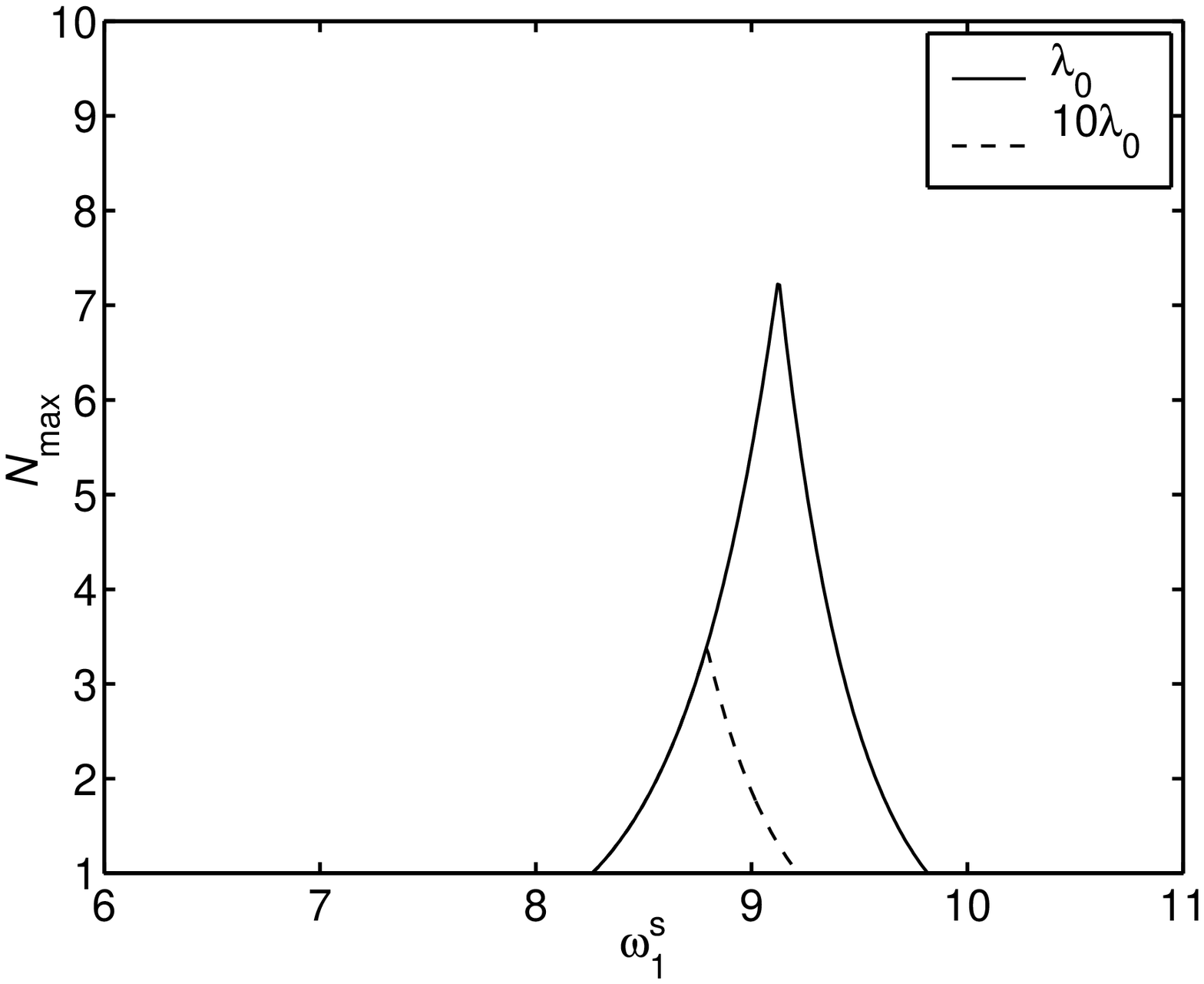,height=200pt}
\caption{Dependence of the maximum number of 20 \AA CdTe nanocrystal
quantum dot qubits for which quantum computation is sustainable, subject to
the three conditions determined by analysis of the two-qubit gate (see
text): i) the fidelity per gate, ${\cal F}={\cal F}_{\max }$, ii) the
antinodal laser intensity, $I_{2}\leq I$ $_{2}^{\max }$, and iii) ${\cal F}
_{\max }>1-\protect\varepsilon ,$ where $\protect\varepsilon $ can be
thought of as the error rate per gate frequency. The figure shows a plot of 
$N_{\rm max}$ as a function of the frequency of the linear support phonon mode, $
\protect\omega _{1}^{s}$, for two values of the linear support densities, $
\protect\lambda _{0}=10$ $\frac{\rm amu}{\text \AA}$, $\protect\lambda =10\protect\lambda
_{0}$ and $\varepsilon=0.1$. The extremum of the functions corresponds to the maximum possible
scalability achievable for 20 \AA ~CdTe nanocrystal qubits. For larger
values of $\protect\omega _{1}^{s}$, the larger values of ${\cal F}^{max}$
which are possible in principle are offset by the need for higher intensity
lasers. In this situation it is possible to support more qubits than are
shown here by relaxing the first constraint. However one thereby loses the
advantage of the increase in ${\cal F}_{\max }$ as $\protect\omega _{1}^{s}$
is increased.}
\label{cdte}
\end{figure}

\begin{figure}[h]
\epsfig{figure=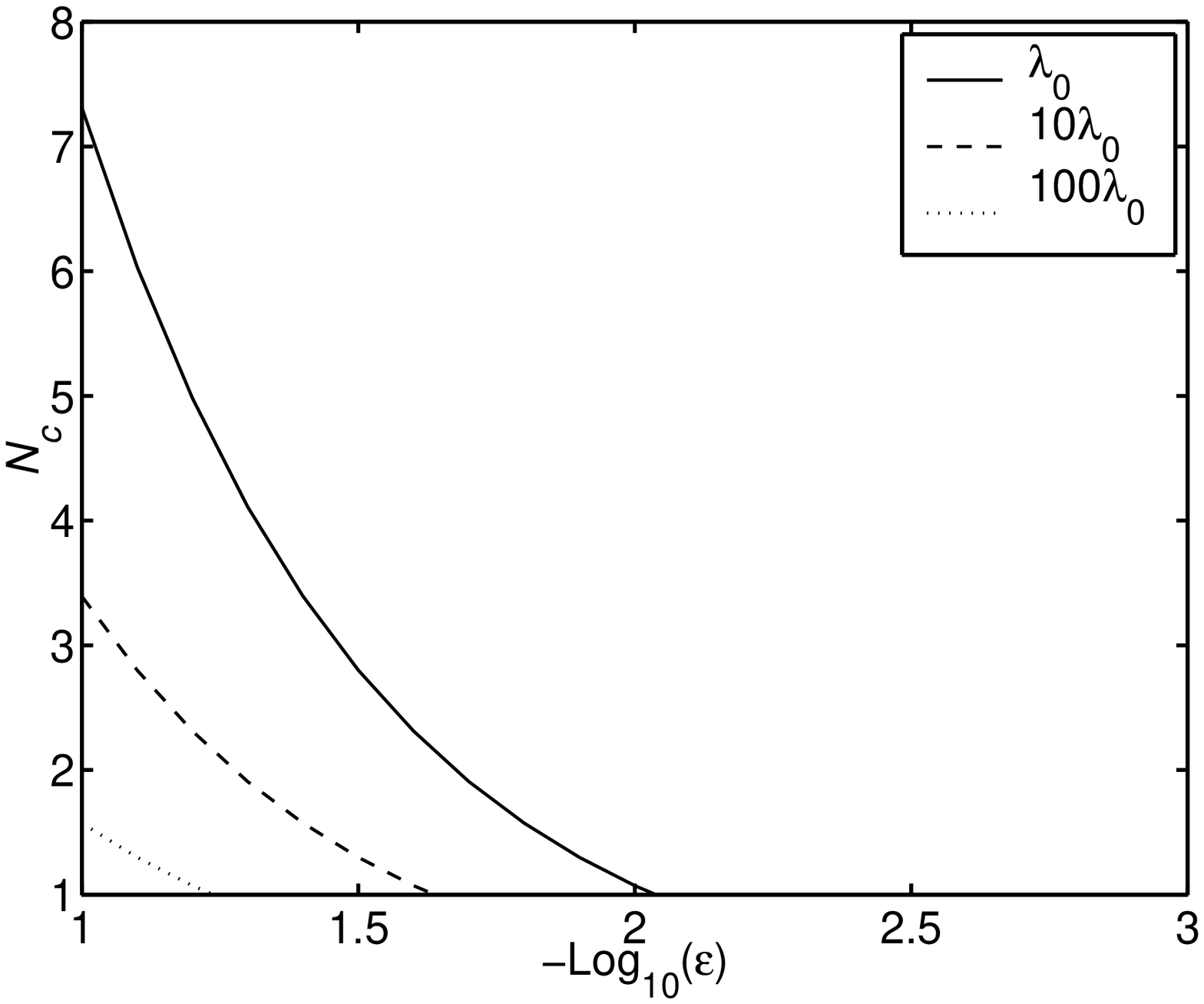,height=200pt}
\caption{Dependence of the optimal number of CdTe nanocrystals, $N_c$, (peaks in Fig \ref{cdte}) on the error threshold $\protect\epsilon$ plotted
for various linear support densities. $\protect\lambda_0 = 10$ $\frac{\rm amu}{\text \AA}$. 
}
\label{cdte2}
\end{figure}

In Figure \ref{cdte}, $N_{\max }$ is plotted as a function of $\omega
_{1}^{s}$ for two linear support densities and\ for a modest threshold of
one error every ten operations ($\varepsilon =0.1$). At low frequencies, Eq (\ref{eq:smallw}) limits $N_{\max }$ and increasing the values of $\omega
_{1}^{s}$ leads to larger values of $N$ for a fixed ${\cal F}_{\max }$. In
contrast, higher frequencies require stronger laser intensities [Eq.(\ref
{eq:intensity_max})] so that eventually the limits on the intensity given in
condition ii) begin to reduce the maximum possible number of quantum dots,
leading to the turnover in Figure~\ref{cdte}. Figure~\ref{cdte} also shows
that the optimal value of $N_{\max }$, which we denote by $N_{c}$ , is
reduced for larger support densities. In Figure~\ref{cdte2}, we now plot $
N_{c}$ as a function of the error threshold $\varepsilon ,$ for a range of
linear densities $\lambda $. We see that even for two qubit quantum devices
one must allow $\varepsilon >0.02,$ or approximately one error every 50
operations. Even at the modest threshold value, $\varepsilon =0.1,$ one
can only support 7 qubits. Clearly, CdTe excitons are thus not a good
candidate for scalable qubits within this scheme. The underlying reason is
that the recombination time of the dark states, while longer than the
operation time, is not sufficiently long to provide high fidelity operations.

\subsection{Si}

Si and other indirect band gap bulk materials exhibit longer exciton
recombination lifetimes than direct band gap materials such as CdTe. 
Although EMA descriptions of Si nanocrystals exist, many subtleties are
required to obtain accurate excitonic states \cite{Hybertsen:94}. These have
also been calculated in semi-empirical tight binding approaches \cite
{KLeung:97}, as well as via pseudopotential methods \cite{Zunger:94}. The
advantage of tight-binding descriptions is that the optical properties of
the nanocrystal can be determined with inclusion of realistic surface
effects \cite{Leung:99}. We estimate the feasibility of using Si
nanocrystals here using the detailed Si excitonic band structure previously
calculated within a semi-empirical description \cite{KLeung:97}. In order
to suppress phonon emission we choose states which correspond to either the
exciton ground state, or lie within the minimal phonon energy of the exciton
ground state. The minimal phonon energies $\omega _{0}^{d}$ are taken from
EMA calculations made by Takagahara \cite{Takagahara:96}, and are
approximately equal to 5 meV for a nanocrystal of 20 \AA\ radius. One
disadvantage of the tight-binding description is that the states are no
longer describable as states with well-defined angular momentum, and the
calculation of electron-phonon coupling is not straightforward. Therefore,
we employ the EMA analysis of Takagahara for this. The Franck-Condon
factors are estimated to be $\sim 0.9$ between electronic states derived
from the same multiplet. Calculations and experiments on Si reveal dark
states with recombination times of microseconds\cite{KLeung:97}. Tight
binding states lack well defined quantum numbers. However, for spherical
dots of 20 \AA\ there are multiple dark states which satisfy our phonon
emission criteria \cite{KLeung:97}. States from these multiplets can be used
to form our logic and auxiliary states. Calculated Raman transitions
between these states have quantitatively similar values to those obtained
for CdTe above. 

\begin{figure}[h]
\epsfig{figure=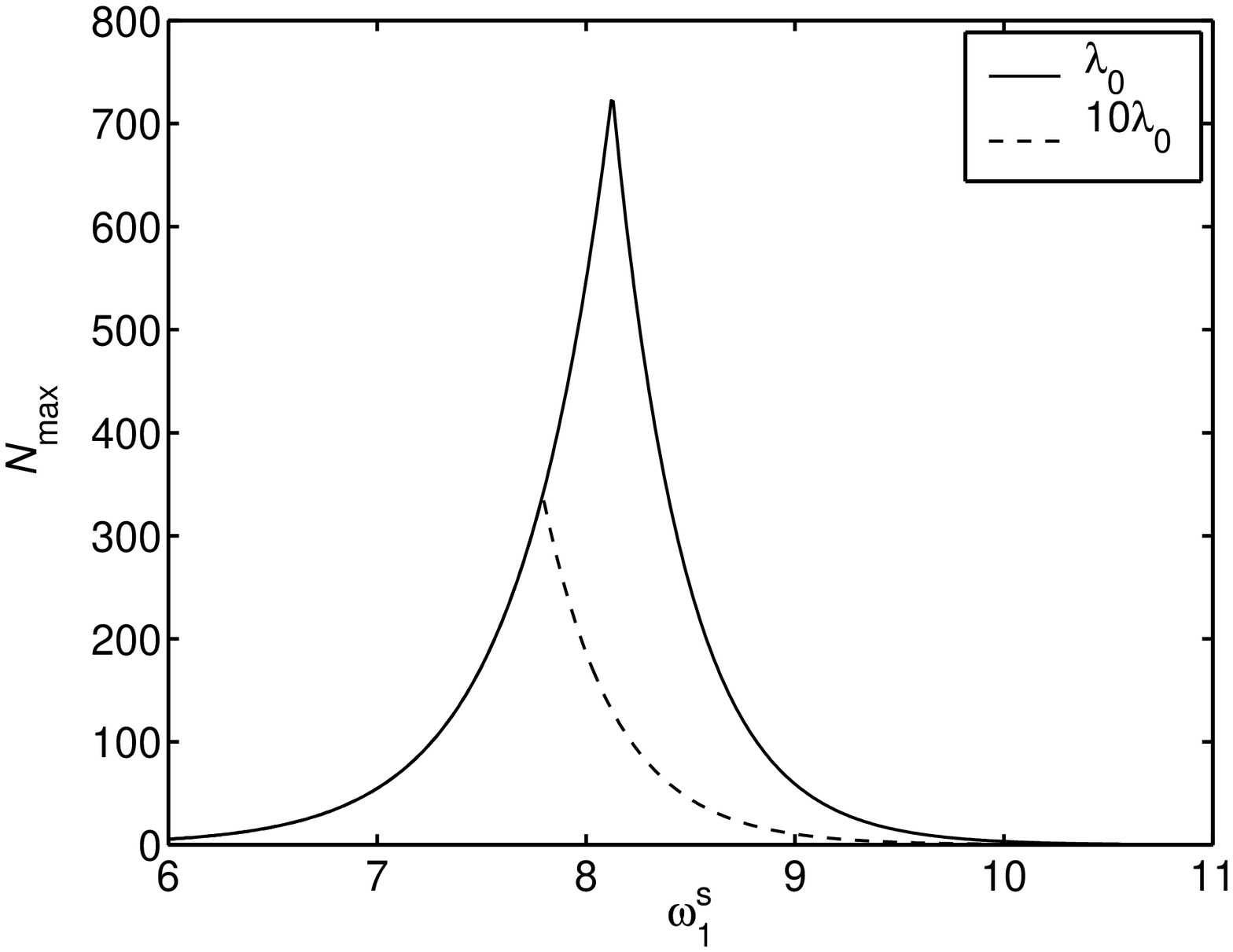,height=200pt}
\caption{Dependence of the maximum number of 20 \AA ~Si nanocrystal quantum
dot qubits for which quantum computation is sustainable, subject to the
three conditions determined by analysis of the two-qubit gate (see text): i)
the fidelity per gate, ${\cal F}={\cal F}_{\max }$,
ii) the antinodal laser intensity, $I_{2}\leq I$ $_{2}^{\max }$, and iii) $
{\cal F}_{\max }>1-\protect\varepsilon ,$ where $\protect\varepsilon $ can
be thought of as the error rate per gate frequency. The figure shows a plot
of $N_{\rm max}$ as a function of the frequency of the linear support phonon
mode, $\protect\omega _{1}^{s}$, for two values of the linear support
densities, $\protect\lambda _{\rm min}=10$ $\frac{\rm amu}{\text \AA}$ and $\protect\lambda =10
\protect\lambda _{\rm min}$ and $\varepsilon=0.1$. The extremum of the functions corresponds to the
maximum possible scalability achievable for 20 \AA ~Si nanocrystal qubits.
The degree of scalability is greater for the indirect band gap material than
for the direct band gap CdTe nanocrystals shown in Figure~\ref{cdte}, and
shows less dependence on the linear support density $\protect\lambda $.}
\label{Si}
\end{figure}

\begin{figure}[h]
\epsfig{figure=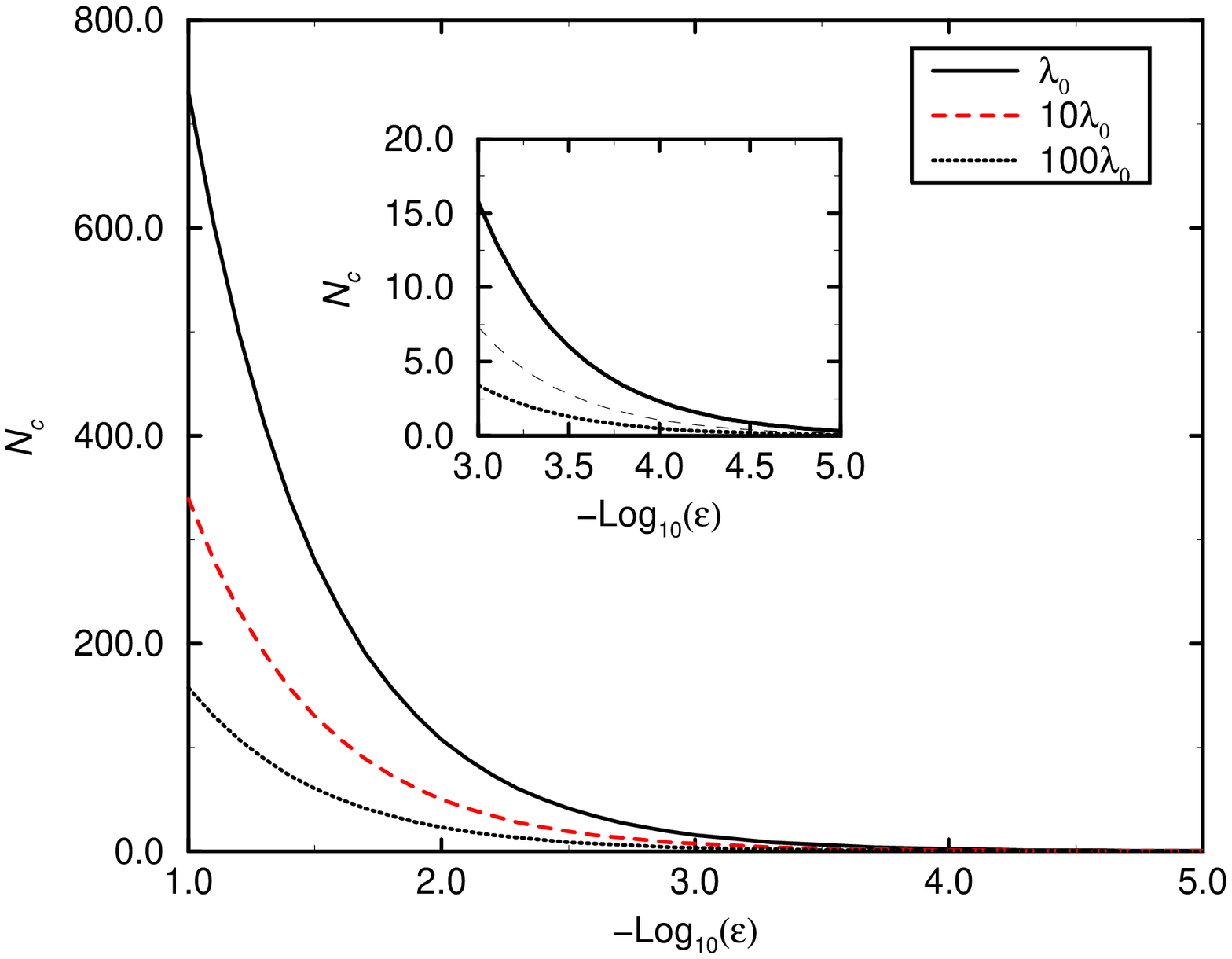,height=200pt}
\caption{Dependence of the optimal numer of Si nanocrystals, $N_c$,
(peaks in Fig \ref{Si}) on the error threshold $\protect\epsilon$
plotted for various linear support densities. $\protect\lambda_0 = 10$
$\frac{\rm amu}{\text \AA}$.}
\label{si2}
\end{figure}

Given an assumed radiative recombination rate $\Gamma =10^{3}$ \cite
{KLeung:97}, we perform an analysis similar to the one above for CdTe.
In Figure \ref{Si} $N_{\max }$ is plotted as a function of $\omega _{1}^{s}
$ for a variety of densities $\lambda $ and a threshold of one error every
10 gates ($\varepsilon =0.1$).
Figure \ref{Si} implies that one could construct a quantum
computer composed of 700 quantum dots if $\varepsilon =0.1$.
In Figure \ref{si2}, the extremum value of $N_{\max}$, $N_{c}$ is plotted as a function
of $\varepsilon $ for a range of $\lambda $ values. The results are also
summarized in Table 1. One sees that, unlike CdTe, for Si there
does now exist the possibility of building a quantum processor that possesses an appreciably lower error rate of $\sim $1 error every
thousand gates. Most encouragingly, it seems possible to construct a
small quantum information processor (5-10 qubits) with a larger linear
support density $10 \lambda_0$, and an error rate of $\varepsilon \leq 10^{-3}$.
Naturally, from an experimental perspective it would probably be more
realistic to use a support having a density at least ten times greater than
our proposed minimal density $\lambda_0$ that was estimated for a pure carbon chain
(e.g., DNA \cite{Alivisatos:96b}, carbon nanotubes, etched supports, etc.).

\section{Conclusions}

\label{conclusions}

We have developed a condensed phase scheme for a quantum computer that is
analogous to the gas phase ion trap proposal and have explored the
feasibility of implementing this scheme with semiconductor quantum dots
coupled by a quantum linear support consisting of a string or rod. We have
found that the Cirac-Zoller scheme of qubits coupled by a quantum phonon
information bus is also applicable in the solid state, and that there exist
some advantages to a condensed phase implementation. One such advantage is
that there is a potential for significantly less noise in the information
bus than in the corresponding gas phase scheme. Calculations by Roukes and
co-workers\cite{Roukes} suggest that much higher $Q$ factors may be found
for nanorods than are currently obtainable in ion traps. Clearly the
extent of the usefulness of our proposal will be very dependent on the
choice of materials. To that end we have analyzed the fidelity for two-qubit
operations for several candidate systems, including both direct and indirect
gap semiconductor quantum dots. We have presented the results of numerical
calculations for implementation of the scheme with CdTe and Si quantum dots,
coupled via either quantum strings or rods. While neither of these
prototypical direct and indirect band gap materials reaches the level of
fidelity and size required for large scale quantum computation, the
indirect gap quantum dots (Si) do show a reasonably high fidelity with an
array of a few tens of dots.

One very revealing result of these explicit calculations of fidelity for
one- and two-qubit gates is the limited scalability. The scheme initially
appears highly scalable in concept due to the solid-state based
architecture. However the detailed analysis given here showed that the
dependence of the Lamb-Dicke parameter $\eta $ on the mass of the
support is a basic problem which essentially limits the
scalability to a few tens of qubits even in the more favorable indirect gap
materials. The main drawback of this condensed phase scheme over the ion
trap scheme is therefore the large reduction in $\eta $ deriving from the
introduction of massive supports. Such a reduction has two important
consequences. First, the laser intensities need to be increasingly large
to perform operations faster than the decoherence time. Second, such large
laser intensities necessitate the use of nodal and antinodal lasers \cite
{James:98,Blatt:2000}. Without these features, the probability of gate
error is extremely high due to transitions to the carrier. This means that
several of the alternative schemes proposed for ion trap computation\cite
{Sorensen:99,Knight:2000} would not provide feasible condensed phase analogs
(although the recent scheme of Childs and Chuang\cite{Childs:00} which
allows computation with two-level ions (or quantum dots) by using both the
blue and red sidebands could also be feasible in the condensed phase).

Additional sources of decoherence which have been neglected here (scattering
off the support, vibrational and electronic transitions in the support) will
also act to limit the number of operations. However one source of
decoherence which can be eliminated or at least reduced, is dephasing from
the coupling to phonon modes of the support. This is a consequence of the
requirement of extremely narrow band-width lasers, and therefore implies that
a similar lack of dephasing will hold for other optical experiments on
quantum dots which use narrow band-widths. One such example is the proposal
to couple quantum dots via whispering gallery modes of glass microspheres 
\cite{Brun:00}. More generally, this result offers a route to avoid
dephasing for other spectral measurements on quantum dots\cite{brown:00b}.

An interesting additional application for this proposal is the laser cooling
of nanorods. A single QD could be placed or even etched on a nanostructure.
A laser tuned to the red support phonon side band of a QD excited
electronic state would excite the energy of the nanocrystal, and at the same
time lower the average phonon occupation of the support. When the unstable
state relaxes, the most probable transition is the carrier transition. The
net result is that the emitted phonon is blue shifted compared to the
excitation pulse. The extra energy carried away by the emitted photon is
thereby removed from the motional energy of the QD.

The essential physical problem encountered in this condensed phase
realization of the qubits coupled by phonon modes is the recombination
lifetime of the qubit states, i.e., the exciton radiative lifetime. In
principle this could be ameliorated by using hyperfine states of a doped
nanocrystal. Recent experimental results demonstrating electronic doping of
semiconductor quantum dots offer a potential route to controlled access of
these states \cite{Shim:00}. The feasibility study presented in this paper
does indicate that although the detailed physics of the qubits and their
coupling is considerably more complicated in the condensed phase than in the
gas phase, limited quantum computation may be possible with phonon-coupled
solid state qubits. Further analysis and development of suitable nanoscale
architectures and materials is therefore warranted.

\section{Acknowledgments}

This work was supported in part by the National Security Agency (NSA) and
Advanced Research and Development Activity (ARDA) under Army Research Office
(ARO) contracts DAAG55-98-1-0371 and DAAD19-00-1-0380. The work of KRB was
also supported by the Fannie and John Hertz Foundation. We 
thank Dave Bacon and Dr. Kevin Leung for useful discussions.

\section{Appendix: Coordinate Representation of Electron and Hole States}

\label{app:Efrosstates}

We give here the coordinate representation of the electron and hole states.
These states were derived in \cite{Efros:96} (We employ a slightly different
phase convention. Efros {\it et al.} when calculating the exchange Hamiltonian, $
\gamma S_{h}\cdot S_{e}$, between the hole and electron spin, $S_{h}$ and $
S_{e}$, use the convention, $S_{h}\cdot S_{e}=S_{h}^{z}S_{e}^{z}+\frac{i}{2}
(S_{h}^{+}S_{e}^{-}-S_{h}^{-}S_{e}^{+})$. We instead use the convention that 
$S_{h}\cdot S_{e}=S_{h}^{z}S_{e}^{z}+\frac{1}{2}
(S_{h}^{+}S_{e}^{-}+S_{h}^{-}S_{e}^{+})$ ). For convenience, we repeat the
definitions of our qubit states [Eq.~(\ref{eq:|1>})] in slightly more
detailed notation:\ 
\begin{eqnarray*}
|0\rangle &=&|\Psi _{1/2,1/2}^{S}({\bf r}_{e})\rangle |\Psi _{3/2,3/2}^{S}(
{\bf r}_{h})\rangle \\
|1\rangle &=&|\Psi _{1/2,-1/2}^{S}({\bf r}_{e})\rangle |\Psi _{3/2,-3/2}^{S}(
{\bf r}_{h})\rangle \\
|2\rangle &=&\frac{1}{\sqrt{2}}\left[ |\Psi _{1/2,-1/2}^{S}({\bf r}
_{e})\rangle |\Psi _{3/2,1/2}^{S}({\bf r}_{h})\rangle -|\Psi _{1/2,1/2}^{S}(
{\bf r}_{e})\rangle |\Psi _{3/2,-1/2}^{S}({\bf r}_{h})\rangle \right] .
\end{eqnarray*}
The electron states are simply the solutions to a free spin-$1/2$ particle
in a spherical hard-wall box: 
\begin{equation}
|S_{e}\rangle \equiv |\Psi _{1/2,\pm 1/2}^{S}({\bf r})\rangle =\sqrt{\frac{2
}{R}}\frac{\sin (\pi r/R)}{r}Y_{0}^{0}(\theta ,\varphi )|S,\pm \frac{1}{2}
\rangle
\end{equation}
where $R$ is the radius of the dot, and $Y_{l}^{m}$ are spherical harmonics. 
$S$ is a conduction band Bloch function and $\pm \frac{1}{2}$ is the $z$
-projection of the electron spin: 
\begin{equation}
|S,\pm \frac{1}{2}\rangle =|L=0,m_{L}=0\rangle |m_{s}=\pm \frac{1}{2}\rangle
.
\end{equation}
The holes states can be written explicitly as: 
\begin{eqnarray*}
|\Psi _{3/2,\pm 1/2}^{S}({\bf r})\rangle &=&-R_{0}(r)Y_{0}^{0}|u_{\pm
1/2}\rangle -R_{2}(r)\left( \sqrt{\frac{2}{5}}Y_{2}^{\pm 2}|u_{\mp
3/2}\rangle +\sqrt{\frac{2}{5}}Y_{2}^{\mp 1}|u_{\pm 3/2}\rangle -\sqrt{\frac{
1}{5}}Y_{2}^{0}|u_{\pm 1/2}\rangle \right) |\Psi _{3/2,\pm 3/2}^{S}({\bf r}
)\rangle \\
&=&-R_{0}(r)Y_{0}^{0}|u_{\pm 3/2}\rangle -R_{2}(r)\left( \sqrt{\frac{2}{5}}
Y_{2}^{\pm 2}|u_{\mp 1/2}\rangle -\sqrt{\frac{2}{5}}Y_{2}^{\pm 1}|u_{\pm
1/2}\rangle +\sqrt{\frac{1}{5}}Y_{2}^{0}|u_{\pm 3/2}\rangle \right)
\end{eqnarray*}
where $R_{l}$ are the envelope functions and $|u_{m_{J}}\rangle $ are the
valence band Bloch functions.

The radial functions are: 
\begin{eqnarray*}
R_{2}(r) &=&\frac{A}{R^{3/2}}\left[ j_{2}(\phi r/R)+\frac{j_{0}(\phi )}{
j_{0}(\phi \sqrt{\beta })}j_{2}(\phi \sqrt{\beta }r/R)\right] \\
R_{0}(r) &=&\frac{A}{R^{3/2}}\left[ j_{0}(\phi r/R)-\frac{j_{0}(\phi )}{
j_{0}(\phi \sqrt{\beta })}j_{0}(\phi \sqrt{\beta }r/R)\right]
\end{eqnarray*}
\newline
where $j_{l}$ are spherical Bessel functions, $\beta =m_{lh}/m_{hh}$ is the
ratio of the light to heavy hole masses, and $\phi $ is the first root of
the equation 
\[
j_{0}(\phi )j_{2}(\sqrt{\beta }\phi )+j_{2}(\phi )j_{0}(\sqrt{\beta }\phi
)=0. 
\]
The constant $A$ is defined by the normalization condition 
\[
\int_{0}^{R}\left[ R_{0}^{2}(r)+R_{2}^{2}(r)\right] r^{2}dr=1. 
\]

The valence band Bloch functions are given by: 
\begin{eqnarray*}
|u_{\pm 3/2}\rangle &=&|L=1,m_{L}=\pm 1\rangle |m_{s}=\pm 1/2\rangle \\
|u_{\pm 1/2}\rangle &=&\frac{1}{\sqrt{3}}\left[ \sqrt{2}|L=1,m_{L}=0\rangle
|m_{s}=\pm 1/2\rangle +|L=1,m_{L}=\pm 1\rangle |m_{s}=\mp 1/2\rangle \right]
.
\end{eqnarray*}
\qquad

For our Raman transitions scheme, we have used states which were not
analyzed in \cite{Efros:96}. In particular, these states are from the $
1S_{e}1P_{5/2}$ exciton multiplet. To find these we used techniques developed
in \cite{Grigoryan:90}, and then calculated the eigenstates of the exchange
coupling using the method of \cite{Efros:96}. A Raman transition connects
states of equal parity through a state of opposite parity. Therefore, the states of interest to us are the $
F_{z}=\pm 1$ state:

\[
|\Psi _{\pm 1}^{v}\rangle =-\frac{1}{\sqrt{3}}\left[ |\Psi _{1/2,\mp
|1/2}^{S}({\bf r}_{e})\rangle |\Psi _{5/2,\pm 3/2}^{P}({\bf r}_{h})\rangle +
\sqrt{2}|\Psi _{1/2,\pm 1/2}^{S}({\bf r}_{e})\rangle |\Psi _{5/2,\pm
1/2}^{P}({\bf r}_{h})\rangle \right] . 
\]
The electron state is as above. The hole state can be written explicitly
as 
\begin{eqnarray*}
|\Psi _{5/2,\pm 3/2}^{P}({\bf r})\rangle &=&R_{1}(r)\left( \sqrt{\frac{2}{5}}
Y_{1}^{0}|u_{\pm 3/2}\rangle +\sqrt{\frac{3}{5}}Y_{1}^{\pm 1}|u_{\pm
1/2}\rangle \right) \\
&&+R_{3}(r)\left( 3\sqrt{\frac{1}{35}}Y_{3}^{0}|u_{3\pm /2}\rangle -\frac{1}{
2}\sqrt{\frac{7}{5}}Y_{3}^{\pm 1}|u_{\pm 1/2}\rangle +\sqrt{\frac{1}{14}}
Y_{3}^{\pm 2}|u_{\mp 1/2}\rangle +\frac{3}{2}\sqrt{\frac{1}{7}}Y_{3}^{\pm
3}|u_{\mp 3/2}\rangle \right)
\end{eqnarray*}
\begin{eqnarray*}
|\Psi _{5/2,\pm 1/2}^{P}({\bf r})\rangle &=&R_{1}(r)\left( \sqrt{\frac{1}{10}
}Y_{1}^{\mp 1}|u_{\pm 3/2}\rangle \sqrt{\frac{3}{5}}Y_{1}^{0}|u_{\pm
1/2}\rangle +\sqrt{\frac{3}{10}}Y_{1}^{\pm 1}|u_{\mp 1/2}\rangle \right) \\
&&+R_{3}(r)\left( 3\sqrt{\frac{3}{70}}Y_{3}^{\mp 1}|u_{\pm 3/2}\rangle -
\sqrt{\frac{6}{35}}Y_{3}^{0}|u_{\pm 1/2}\rangle -\sqrt{\frac{1}{70}}
Y_{3}^{\pm 1}|u_{\mp 1/2}\rangle +\sqrt{\frac{3}{7}}Y_{3}^{\pm 2}|u_{\mp
3/2}\rangle \right) ,
\end{eqnarray*}
where $R_{l}$ are the envelope functions and $|u_{m_{J}}\rangle $ are the
valence band Bloch functions given above.

The radial functions are 
\begin{eqnarray*}
R_{3}(r) &=&\frac{B}{R^{3/2}}\left[ j_{3}(\phi ^{\prime }r/R)+\frac{
2j_{1}(\phi ^{\prime })}{3j_{1}(\phi ^{\prime }\sqrt{\beta })}j_{3}(\phi
^{\prime }\sqrt{\beta }r/R)\right] \\
R_{1}(r) &=&\frac{B}{R^{3/2}}\left[ j_{1}(\phi ^{\prime }r/R)-\frac{
j_{1}(\phi ^{\prime })}{j_{1}(\phi ^{\prime }\sqrt{\beta })}j_{1}(\phi
^{\prime }\sqrt{\beta }r/R)\right] ,
\end{eqnarray*}
where $\phi ^{\prime }$ is the first root of the equation 
\begin{equation}
j_{1}(\phi )j_{3}(\sqrt{\beta }\phi )+\frac{2}{3}j_{1}(\phi )j_{3}(\sqrt{
\beta }\phi )=0
\end{equation}
and $B$ is defined by the normalization condition 
\begin{equation}
\int_{0}^{R}\left( R_{1}^{2}(r)+R_{3}^{2}(r)\right) r^{2}dr=1.
\end{equation}

\bigskip

Table 1. For a given error threshold, $\varepsilon$, and support density $
\lambda$, the table shows the optimal value of $N_{\max },N_{c}$ for CdTe
and Si nanocrystals. $\lambda_0=10$ mu/\AA .\label{tablea}

\bigskip 
\begin{tabular}{cccc}
-$\log _{10}$($\varepsilon $) & $\lambda/\lambda_0 $ & $N_{c}$(CdTe) & $N_{c}$(Si) \\ 
\hline
1 & 1 & 7 & 731 \\ 
& 10 & 3 & 339 \\ 
& 100 & 1 & 158 \\ 
2 & 1 & 1 & 107 \\ 
& 10 & 0 & 50 \\ 
& 100 & 0 & 23 \\ 
3 & 1 & 0 & 16 \\ 
& 10 & 0 & 7
\end{tabular}
\bigskip

\end{document}